\documentclass[12pt]{iopart}
\usepackage{graphicx}
\usepackage{xcolor}
\usepackage{color}

\newcommand\Nv{{\mathbf{N}}}
\def \beq {\begin{equation}}
\def \eeq {\end{equation}}
\def \bea {\begin{eqnarray}}
\def \eea {\end{eqnarray}}
\def \bfig {\begin{figure}}
\def \efig {\end{figure}}

\def \lab {\label}

\def \bB {\textbf{B}}

\def \be {\textbf{e}}

\begin{document}

\title[]{Differential kinetic dynamics and heating of ions in the turbulent solar wind}

\author{F. Valentini$^1$, D. Perrone$^2$, S. Stabile$^1$, O. Pezzi$^1$, S. Servidio$^1$, R. De Marco$^3$, F. Marcucci$^3$, R.
Bruno$^3$, B. Lavraud$^{4,5}$, J. De Keyser$^6$, G. Consolini$^3$, D. Brienza$^3$, L. Sorriso-Valvo$^7$, A. Retin\`o$^8$, A.
Vaivads$^9$, M. Salatti$^{10}$, P. Veltri$^1$} 
\address{$^1$ Dipartimento di Fisica, Universit\`a della Calabria, Rende (CS), I-87036, Italy}
\address{$^2$ European Space Agency, Science and Robotic Exploration Directorate, ESAC, Madrid, Spain}
\address{$^3$ INAF-IAPS Istituto di Astrofisica e Planetologia Spaziali, 00133 Roma, Italy}
\address{$^4$ Institut de Recherche en Astrophysique et Plan\'etologie, Université de Toulouse, France}
\address{$^5$ Centre National de la Rescherche Scientifique, UMR 5277, Toulouse, France}
\address{$^6$ Royal Belgian Institute for Space Aeronomy, B-1180, Brussels, Belgium }
\address{$^7$ CNR-Nanotec, Rende (CS), I-87036, Italy}
\address{$^8$ Laboratoire de Physique des Plasmas, Ecole Polytechnique, Palaiseau, F-91128, France}
\address{$^9$ Swedish Institute of Space Physics, Uppsala, 75121, Sweden}
\address{$^{10}$ Agenzia Spaziale Italiana, 00133 Roma, Italy}
\ead{francesco.valentini@fis.unical.it}
\vspace{10pt}
\begin{indented}
\item[]February 2014
\end{indented}

\begin{abstract}
The solar wind plasma is a fully ionized and turbulent gas ejected by the outer layers of the solar corona at very high speed,
mainly composed by protons and electrons, with a small percentage of helium nuclei and a significantly lower abundance of heavier
ions. Since particle collisions are practically negligible, the solar wind is typically not in a state of thermodynamic
equilibrium. Such a complex system must be described through self-consistent and fully nonlinear models, taking into account its
multi-species composition and turbulence. We use a kinetic hybrid Vlasov-Maxwell numerical code to reproduce the turbulent energy
cascade down to ion kinetic scales, in typical conditions of the uncontaminated solar wind plasma, with the aim of exploring the
differential kinetic dynamics of the dominant ion species, namely protons and alpha particles. We show that the response of
different species to the fluctuating electromagnetic fields is different. In particular, a significant differential heating 
of alphas with respect to protons is observed. Interestingly, the preferential heating process occurs in spatial regions nearby
the peaks of ion vorticity and where strong deviations from thermodynamic equilibrium are recovered. Moreover, by feeding a
simulator of a top-hat ion spectrometer with the output of the kinetic simulations, we show that measurements by such 
spectrometer planned on board the Turbulence Heating ObserveR (THOR mission), a candidate for the next M4 space mission of the
European Space Agency, can provide detailed three-dimensional ion velocity distributions, highlighting important non-Maxwellian
features. These results support the idea that future space missions will allow a deeper understanding of the physics of the
interplanetary medium.
\end{abstract}

\section{Introduction}
The solar wind is a rarefied, highly variable flow of charged particles originating from the Sun. This medium is composed mainly
by electrons and protons, with a small fraction of alpha particles (helium nuclei) and a lower percentage of heavier ions. Due to
its low particle density, collisions are essentially negligible, and the solar wind plasma is in a state of non-thermodynamic
equilibrium \cite{mar06}. One of the most puzzling aspects of the solar wind dynamics consists in the empirical evidence
\cite{ric03} that it is hotter than expected from adiabatic expansion \cite{gaz82,ric03,bor14}. Understanding the mechanisms of
energy dissipation and particle heating in such a collisionless system represents a real challenge for space plasma physics.

`In situ' spacecraft measurements reveal that the solar wind is in a state of fully-developed turbulence \cite{bru05,bru13}.
The energy injected by large scale solar dynamical features into the Heliosphere, in the form of long-wavelength fluctuations,
cascades towards small scales via nonlinear interactions until it can be transferred to the plasma as heat. In the inertial range,
the power spectrum of the solar wind fluctuations manifests a behavior reminiscent of the Kolmogorov phenomenology for fluid
turbulence \cite{kol41,col68,dob80,tu95,gol95}. The turbulent cascade extends to smaller spatial scales, down to a range of
wavelengths where kinetic effects start playing a non-negligible role. At the typical ion characteristic scales (Larmor radius
and/or inertial length), different physical processes come into play, leading to changes in the spectral shape
\cite{lea98,lea00,mak01,bal05,smi06,ale07,mat07,bou12}. Here, the dynamics of the solar wind ions gets extremely complicated,
being dominated by complex kinetic processes such as resonant wave-particle interactions, particle heating, generation of
temperature anisotropy, production of beams of accelerated particles, etc. In this range of scales the particle velocity
distribution is generally observed to deviate from the typical Maxwellian shape of thermodynamic equilibrium.

Several models have been developed to understand the kinetic scale phenomenology, focusing specifically on  heating and
acceleration processes. To explain the above phenomena, several physical mechanisms have been proposed. Among them, the
ion-cyclotron resonance can produce heating \cite{mat99,oug01,dim01,hol02,cra03,cra05,cra07,iazzolino10}, where ions interact
resonantly with ion-cyclotron waves generated along the cascade. An alternative model is the non-resonant stochastic heating
\cite{cha10,bou13}, where the fluctuations at scales comparable with the ion gyro-scales produce significant distortions of the
ion orbits that become stochastic in the plane perpendicular to the magnetic field, violating the magnetic moment conservation.
Finally, it has been clearly observed that the interaction of particles with coherent structures can locally produce heating and
non-Maxwellian features \cite{gre09,osm11,osm12,ser12,gre12,wu13,Mat15}.

One of the most relevant aspects of the ion heating process in the solar wind is that the heavy ions are preferentially heated and
accelerated with respect to protons. Measurements of solar wind hydrogen and helium temperatures \cite{kas08} show compelling
evidence of Alfv\'en-cyclotron dissipation mechanisms. A more recent observational study \cite{tra16} shows that for the
heavy ion component ($A>4$ amu) the temperature displays a clear dependence on mass, probably reflecting the physical conditions
in the solar corona. Again, the interpretation of these observations is mostly based on the physical processes of wave-particle
interactions \cite{ise09,per11,val11,cra12,per14a,araneda09,maneva13,ofman14,maneva151,maneva152}, stochastic ion heating
\cite{cha13}, and interaction between particles and coherent structures \cite{per13,per14b}. Despite a significant theoretical
effort, there is still no definitive solution to the problem of heavy ions differential heating. It is important to note that the
physical processes of the differential heating of heavy ions are also important for laboratory plasmas research, aiming at heating
a confined plasma to initiate fusion reactions.

In this paper, we study the kinetic dynamics of protons and alpha particles in typical conditions of the interplanetary medium, by
employing multi-component Eulerian hybrid Vlasov-Maxwell (HVM) simulations \cite{val07,per11}. The multicomponent version of the
HVM code integrates the Vlasov equation numerically for the distribution function of both protons and alpha particles, while
treating electrons as a massless, isothermal fluid. We present the numerical results of HVM simulations of decaying turbulence
with guide field, in a 2D-3V phase space domain (two dimensions in physical space and three dimensions in velocity space). 
During the turbulent cascade, the ion species depart locally from the condition of thermodynamic equilibrium
\cite{ser12,per13,val14,ser14,ser15}, exhibiting temperature anisotropy, differential kinetic behavior as well as preferential
heavy ion heating. Interestingly, the preferential heating process occurs in thin filaments of the typical size of few inertial 
lengths, preferentially located in the proximity of vorticity structures. In the same regions, significant deviations from
Maxwellian distributions are observed for ion species. Moreover, as we will discuss in detail below, we have analyzed the
distribution of the numerical data in the fire-hose and mirror stability plane, as dependent on the multi-ion composition of the
interplanetary medium, showing a good agreement between the numerical results and the observational evidences discussed in a
recent paper by Chen et al. \cite{chen16}.
\bfig
  \centering 
\includegraphics[width=8cm]{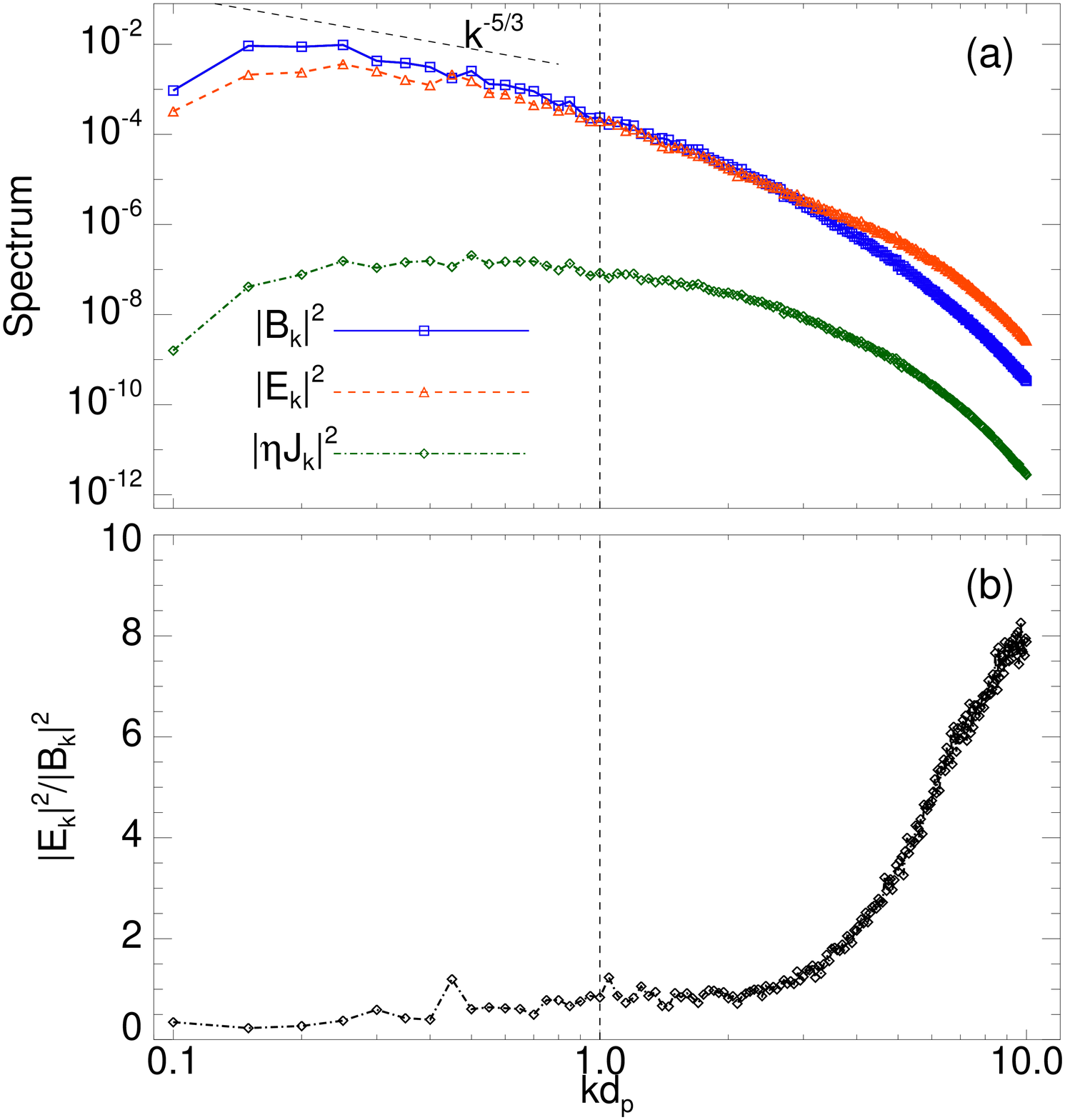}
\caption{Panel (a): Omni-directional spectra of magnetic (blue solid-square line) and electric (red dashed-triangle line)
energy at the maximum of the turbulent activity. Panel (b): Semi-logarithmic plot of the ratio between electric and magnetic
energy. The vertical black dashed lines indicate the proton skin depth characteristic wavenumber. The Kolmogorov expectation,
$k^{-5/3}$, is reported as a black dashed line, while the green diamond curve represents the contribution $|\eta J_k|^2$ to
the electric energy, due to the resistive term in the Ohm's law in Eq. (\ref{eq:ohm}).}
\lab{fig:spec_prot}
\efig	

Finally, through a top-hat simulator \cite{demarco16} using the output of the HVM simulations, we show that the Cold Solar
Wind (CSW) instrument \cite{lavraud16} on board the Turbulence Heating ObserveR - THOR spacecraft \cite{vaivads16}, a candidate
for the next M4 space mission of the European Space Agency, is able to provide adequate high resolution measurements of the ion
velocity distributions for the study of turbulent ion heating. In this regard, the THOR mission will allow to measure the velocity
distributions of the solar wind protons and alpha particles, with unprecedented phase space resolution.

\section{Setup of multi-ion hybrid Vlasov-Maxwell simulations}
To perform a numerical study of the microphysics of a multi-ion turbulent plasma, we employ an Eulerian hybrid Vlasov-Maxwell
(HVM) numerical code \cite{val07,per13} which solves the Vlasov equations for the distribution functions of protons and alpha
particles. In this model electrons are treated as an isothermal massless fluid. Vlasov equations for ions are coupled to a
generalized Ohm's law, in which both the Hall term and the electron pressure contribution are retained, and to the Amp\'ere and
Faraday equations. The displacement current is neglected and quasi-neutrality is assumed. The dimensionless HVM equations can be
written as:
\bea
\label{eq:vlas}
\frac{\partial f_s}{\partial t} + \textbf{v} \cdot \frac{\partial f_s}{\partial \textbf{r}} + \xi_s \left( \textbf{E}+ \textbf{v}
\times \textbf{B} \right) \cdot \frac{\partial f_s}{\partial \textbf{v}}=0, 
\label{eq.1} \\
\label{eq:ohm}
\textbf{E} = - \left( \textbf{u}_e \times \textbf{B} \right) - \frac{1}{n_e} \nabla P_e + \eta \textbf{j},  \\
\label{eq:induction}
\frac{\partial \textbf{B}}{\partial t} = - \nabla \times \textbf{E}. 
\eea
In the above equations, masses and charges have been scaled by the proton mass $m_p$ and charge $e$ respectively, time by the
inverse proton-cyclotron frequency ($\Omega_{cp}^{-1}=m_pc/eB_0$, $B_0$ being the ambient magnetic field and $c$ the speed of
light), velocity by the Alfv\'en speed ($V_A = B_0/\sqrt{4 \pi n_{0,p} m_p}$, $n_{0,p}$ being the equilibrium proton density), in
which we only considered the contribution of the dominant proton species, and length by the proton skin depth ($d_p = V_A /
\Omega_{cp}$). In Eq. (\ref{eq:vlas}), $f_s(\textbf{r},\textbf{v},t)$ is the ion distribution function (the subscript $s = p,
\alpha$ refers to protons and alpha particles, respectively), $\textbf{E}(\textbf{r},t)$ and $\textbf{B}(\textbf{r},t)$ are the
electric and magnetic fields, and $\xi_s$ is a constant that depends on the charge to mass ratio of each ion species ($\xi_p=1$
and $\xi_\alpha=1/2$). In Eq. (\ref{eq:ohm}), the electron bulk velocity $\textbf{u}_e$ is defined as $(\Sigma_s Z_s n_s
\textbf{u}_s - \textbf{j})/n_e$, where $Z_s$ is the ion charge number ($Z_p=1$ and $Z_\alpha=2$), the density $n_s$ and the ion
bulk velocity $\textbf{u}_s$ are zero-th and first order velocity moment of the ion distribution function, respectively. The
electron density is derived from the quasi-neutrality equation $n_e = \Sigma_s Z_s n_s$. Furthermore, $\textbf{j} = \nabla \times
\bB$ is the total current density and $P_e=n_eT_e$ is the electron pressure ($T_e$ is assumed to be constant in time and space). A
small resistive term has been added as a standard numerical Laplacian dissipation ($\eta = 2 \times 10^{-2}$), in order to remove
any spurious numerical effects due to the generation of strong magnetic field gradients during the development of turbulence.

We solve the multi-ion hybrid Vlasov-Maxwell equations in a 2D-3V phase space domain. The system size in the spatial domain is $L
= 2\pi \times 20d_p$ in both $x$ and $y$ directions, where periodic boundary conditions have been implemented, while the limits of
the velocity domain are fixed at $v_{max,s} = \pm 5 v_{th,s}$ in each velocity direction, $v_{th,s}=(k_BT_{0,s}/m_s)^{1/2}$ being
the ion thermal speed at equilibrium and $T_{0,s}$ the equilibrium temperature. In each direction in the velocity domain we set
$f_s(|v|>v_{max,s})=0$. The 5D numerical box has been discretized by $512^2$ grid-points in the 2D spatial domain and $71^3$
grid-points in both proton and alpha-particle 3D velocity domains. A higher velocity resolution has been adopted here with respect
to previous works \cite{per13,per14b}, in order to ensure a significantly improved conservation of the Vlasov invariants and 
to provide a better description of the details of the ion distribution function in velocity space. The time step, $\Delta t$,
has been chosen in such a way that the Courant-Friedrichs-Lewy condition for the numerical stability of the Vlasov algorithm is
satisfied \cite{pey86}. In order to control numerical accuracy, a set of conservation laws is monitored during the simulations:
typical relative variations of the total energy is $\simeq 0.6\%$, of the entropy is $\simeq 0.4\%$ and of the mass is $\simeq 2.5
\times 10^{-3}\%$ for protons and $\simeq 0.35\%$ for alpha particles. 

At $t=0$, both ion species have Maxwellian velocity distributions and homogeneous constant densities. To simulate physical
conditions close to those of the pristine solar wind, at equilibrium we set the alpha particle to proton density ratio $n_{0,
\alpha}/n_{0,p} = 5 \%$ and temperature of each species such that $T_{0,\alpha}=T_{0,p} = T_e$. The plasma is embedded in a
uniform background out of plane magnetic field, $\bB_0$ ($\textbf{B}_0= B_0 \widehat{\bf{e}}_z$). We perturb the initial
equilibrium with a 2D spectrum of Fourier modes, for both proton bulk velocity and magnetic field. The energy is injected with
random phases and wave numbers in the range $0.1 < k < 0.3$, where $k = 2 \pi m/L$, with $2 \leq m \leq 6$. Neither density
disturbances nor parallel variances are imposed on the initial equilibrium, namely $\delta n = \delta u_z = \delta B_z = 0$. We
performed three different simulations whose relevant parameters are summarized in Table \ref{tab}. Below, we will discuss the
results of RUN 1. RUN 2 and RUN 3 are qualitatively similar to RUN 1 and have been considered to have significant statistics to
perform a direct comparison with solar wind data.
\begin{table}
\begin{center}
\begin{tabular} {|c|c|c|c|}
\hline
RUN  & $\beta_p=2v_{th,p}^2/V_A^2$  & $\delta B/B_0$ & $t^*\Omega_{cp}$\\
\hline
1 & 0.5 & 1/3 & 49\\
\hline
2 & 2.0   & 1/3 & 51\\
\hline
3 & 1.0   & 2/3 & 25\\
\hline
\end{tabular}
\end{center}
\caption{Simulation parameters.}
\label{tab}
\end{table}

\section{Turbulence cascade and generation of temperature anisotropy}
We numerically investigate the kinetic evolution of protons and alpha particles in a situation of decaying turbulence. The large
scale fluctuations imposed on the initial equilibrium produce a turbulent cascade down to kinetic scales. In analogy with fluid
models, by looking at the time evolution of the spatially averaged mean squared out of plane current density $\langle j_z^2
\rangle$, it is possible to identify an instant of time $t^*$ at which the turbulent activity reaches its maximum level
\cite{min09}. The value of $t^*$ in units of $\Omega_{cp}^{-1}$ is reported in the right column of Table \ref{tab}, for each
simulation. At this time $t^*$ we perform our analysis \cite{ser12,per13,val14,ser15}.

In panel (a) of Figure~\ref{fig:spec_prot}, we show the omni-directional magnetic $|B_k|^2$ (blue solid-square line) and
electric ${|E_k|^2}$ (red dashed-triangle line)  spectra at $t=t^*$ for RUN 1. The Kolmogorov expectation, $k^{-5/3}$ (black
dashed line), is reported as a reference, while the vertical black dashed line indicates the proton skin depth characteristic
wavenumber. In the range of small wavenumbers the magnetic activity is dominant, while the situation clearly changes for $k
d_p>2$, where the electric energy becomes larger than the magnetic one. To highlight this point, in panel (b) of Figure
\ref{fig:spec_prot} we report the electric to magnetic energy ratio as a function of the wavenumber. The break point between the
two regimes is located around the proton skin depth, where dispersive Hall effects come into play \cite{val14}. These results
present several similarities to spacecraft measurements in the solar wind \cite{bal05,sahraoui09,valentini10}.

As already noticed in previous works \cite{ser12,per13,val14,ser15,per14b}, ion temperature anisotropy is generated along the
development of the turbulent cascade. We focus, at first, on the temperature anisotropy with respect to the direction of the local
magnetic field $T_{\perp}/T_{\parallel}$ for each ion species. The temperature of each ion species has been computed as the
second-order velocity moment of the ion distribution function:
\beq
T_s = \frac{m_s}{3n_s} \int (\textbf{v} - \textbf{u}_s)^2 f_s d^3 v \ \ \ \ \ \ s=p,\alpha .
\eeq 
In Figure \ref{fig1}, we show the 2D contours of the temperature anisotropy of protons (left) and alpha particles (right) at
$t=t^*$, for RUN 1 (the same behavior is observed for RUN 2 and RUN 3); as can be noticed from the comparison of both panels of
this figure, the global bi-dimensional patterns of the temperature anisotropy display similar features, with larger values of
anisotropy being concentrated in thin filaments (of the typical size of few ion skin depths) for both species. However, a
significant differential behavior is observed, as the alpha particles appear consistently more anisotropic than the protons. 
\begin{figure*}
\centering 
\includegraphics[width=7.5cm]{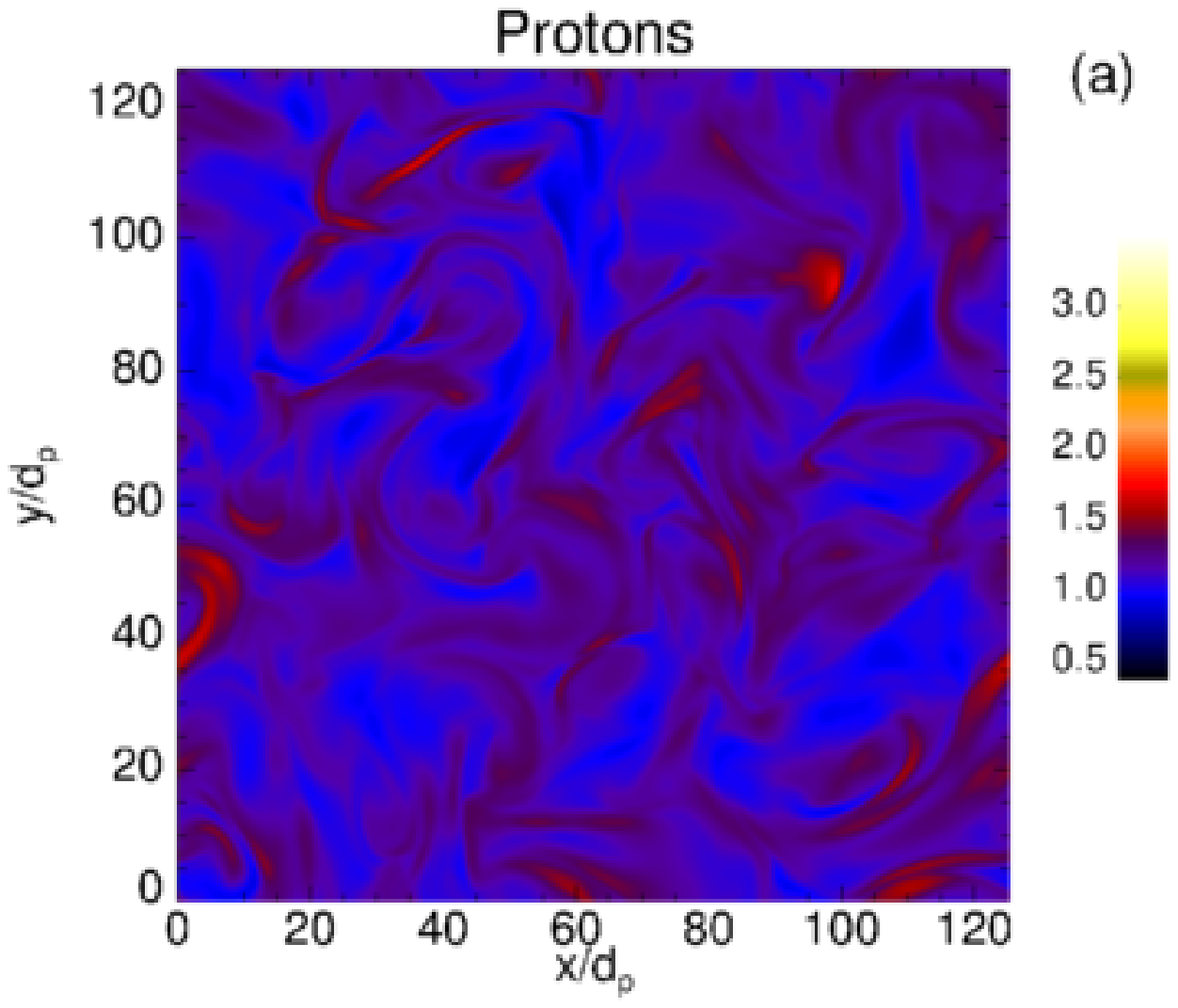}
\includegraphics[width=7.5cm]{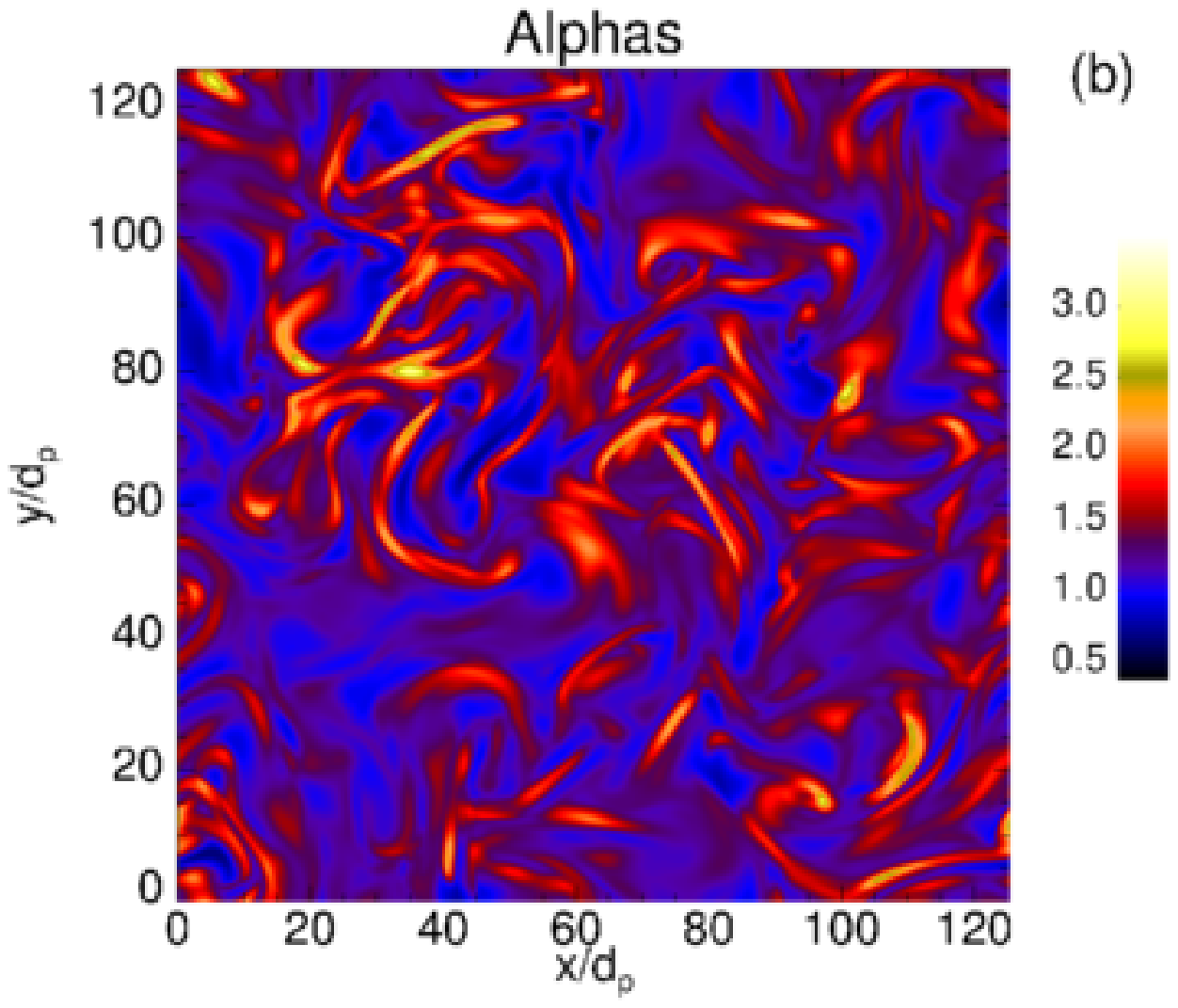}
\caption{2D contour plot of the temperature anisotropy $T_\perp/T_\parallel$ with respect to the direction of the local magnetic
field for protons (a) and alpha particles (b).}
\lab{fig1}
\end{figure*}

With the aim of showing that the numerical results of the ensemble of simulations in Table \ref{tab} reproduce closely the typical
behavior observed in the solar wind multi-ion plasma, we consider here a recent paper by Chen et al. \cite{chen16}, in which
solar wind ion temperature anisotropy from the {\it Wind} spacecraft at 1 AU has been analyzed. The authors proposed a detailed
study of the plasma stability, as dependent on the simultaneous presence of protons, electrons and alpha particles, as main
constituents of the solar wind plasma. Specifically, they made a comprehensive analysis based on three years of data and gave
clear evidence that the fire-hose and mirror instability thresholds, calculated in Refs. \cite{kuntz15,hellinger07} for a
multi-species plasma system, well constrain the whole data distribution to the stable side and that the contours of the
distribution follow the shape of the thresholds.

\begin{figure}
\centering 
\includegraphics[width=7.5cm]{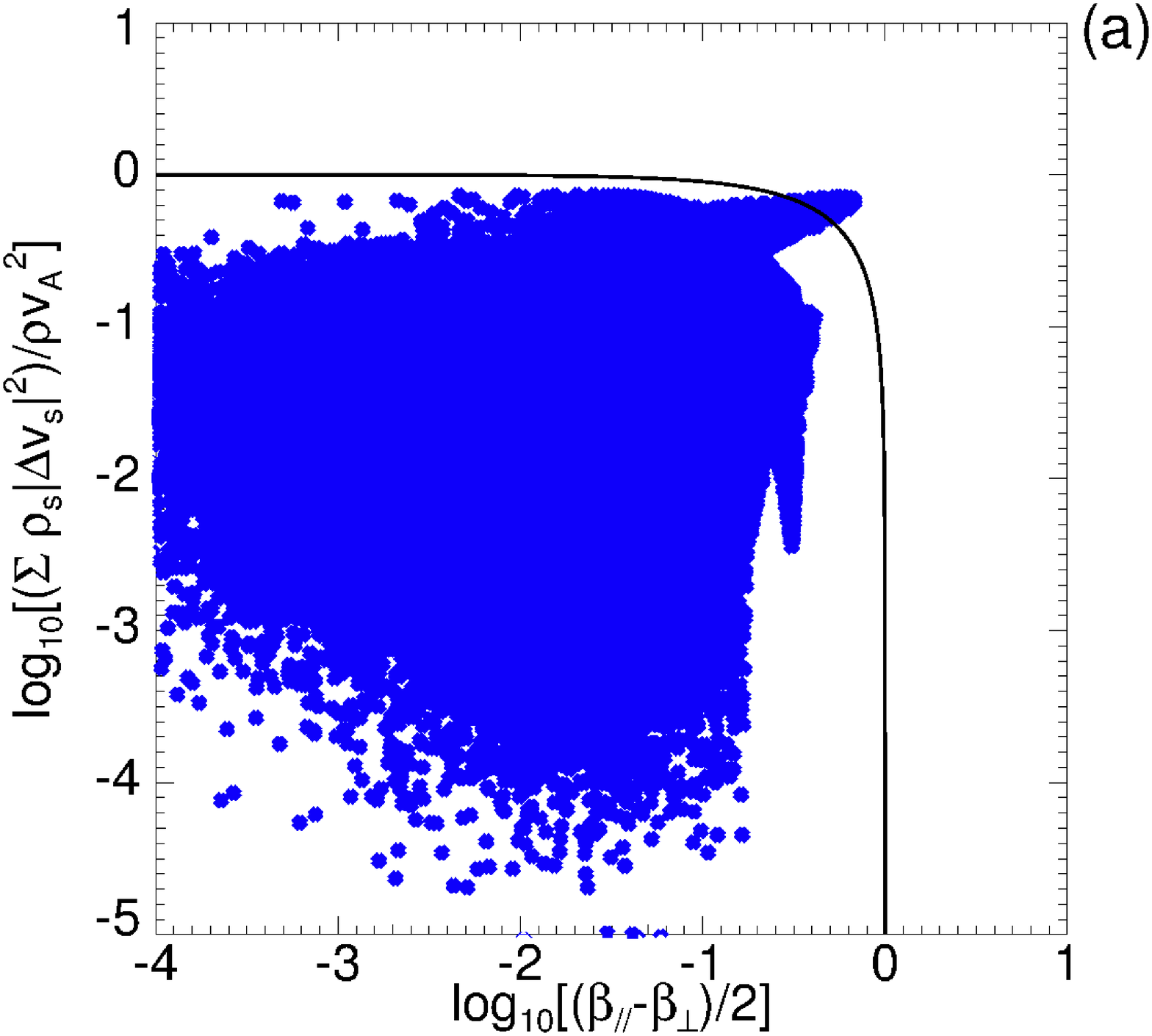}
\includegraphics[width=7.5cm]{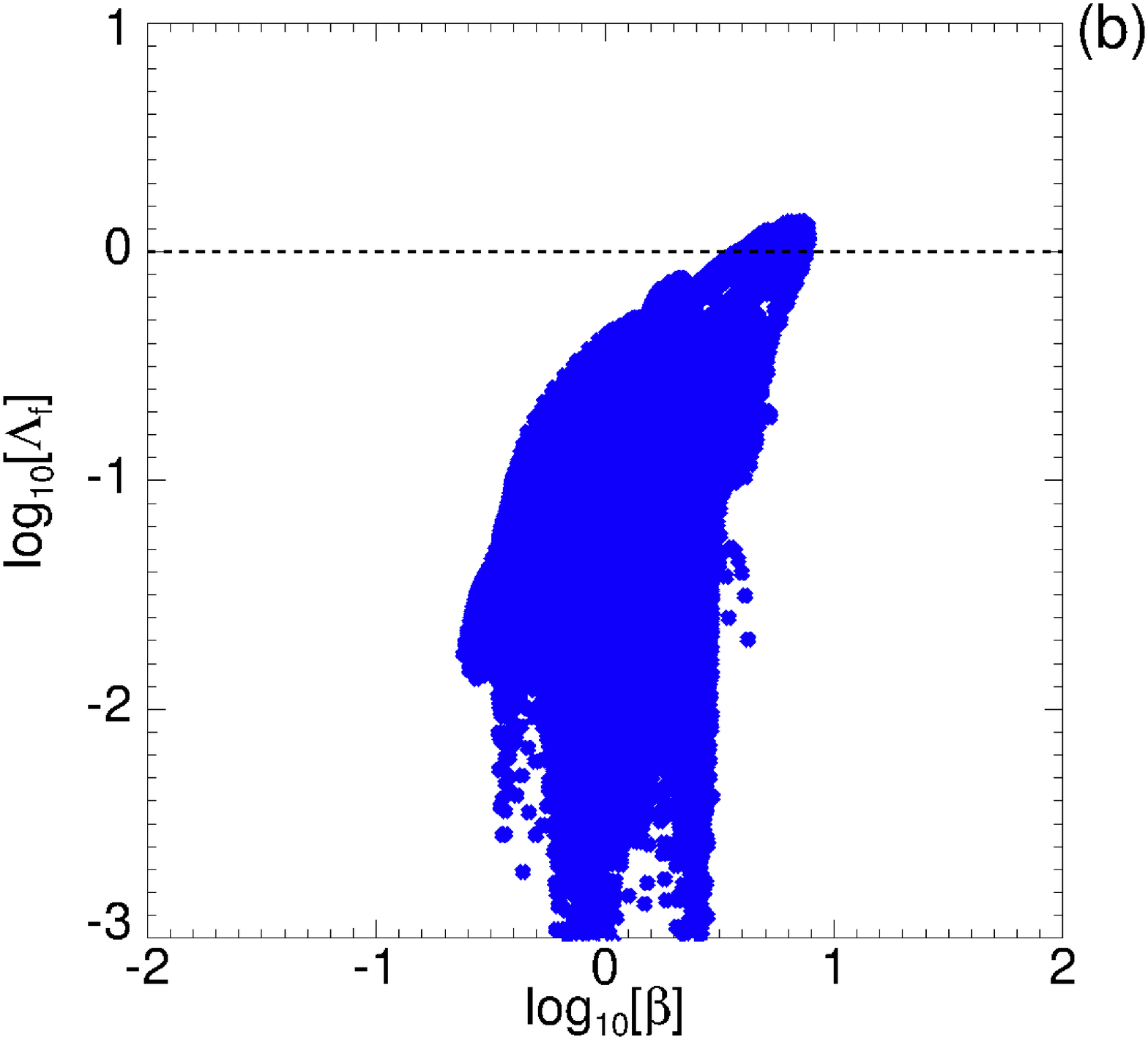}\\
\includegraphics[width=7.5cm]{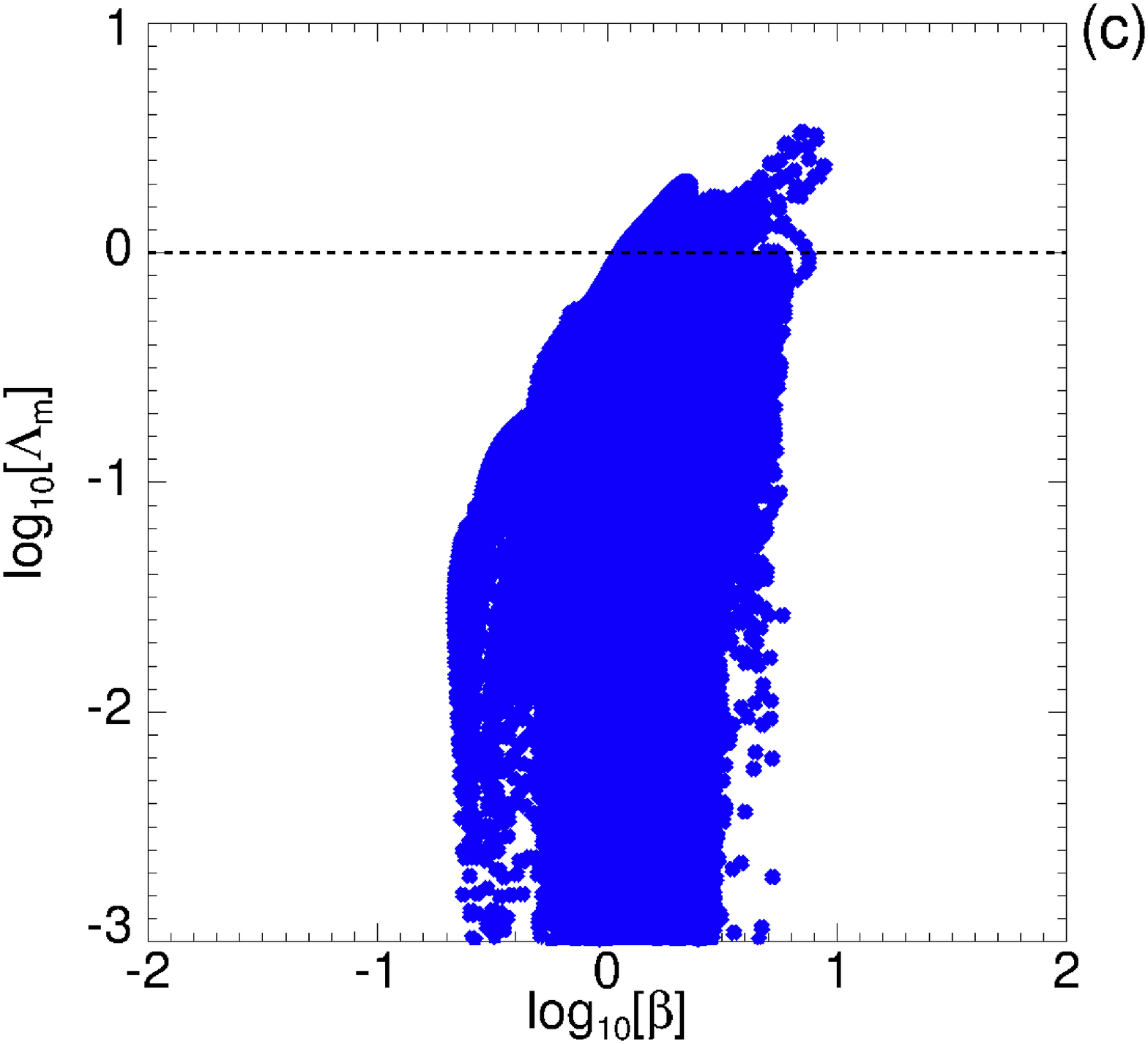}
\caption{(a) Scatter plot of the  two terms of $\Lambda_f$ for the ensemble of simulations in Table \ref{tab}. Solid (black) line
represent the instability threshold. Panels (b) and (c): $\log_{10}\Lambda_f$ and $\log_{10}\Lambda_m$, respectively, as a
function of $\log_{10}\beta$.}
\lab{fig2}
\end{figure}
In order to compare with these observations, we have considered the conditions for the long-wavelength fire-hose and mirror
instabilities in a multi-species plasma \cite{kuntz15,hellinger07}:
\begin{eqnarray}
 & &\Lambda_f=\frac{\beta_\parallel-\beta_\perp}{2}+\frac{\sum_s \rho_s|\Delta {\bf v}_s|^2}{\rho v_A^2}>1, \\
 & &\Lambda_m=\sum_s\beta_{\perp s}\left(\frac{\beta_{\perp s}}{\beta_{\parallel s}}-1\right)-\frac{\left(\sum_s
q_sn_s\beta_{\perp s}/\beta_{\parallel s}\right)^2}{2\sum_s (q_sn_s)^2/\beta_{\parallel s}}>1,
\end{eqnarray}
where $s$ stands for protons, alphas and electrons respectively ($s=p,\alpha,e$),
$\beta_{\perp,\parallel}=\sum_s\beta_{\perp,\parallel s}$, $\rho_s$ is the mass density of the species $s$, $\rho$ is the total
mass density, $v_A$ is the local Alfv\'en speed, $\Delta {\bf v}_s$ is the difference between the bulk velocity of the species $s$
and the center of mass velocity. In Figure \ref{fig2} [panel (a)] we report a scatter plot of the first and second term of
$\Lambda_f$ (logarithmic axes), for all the cases in Table \ref{tab}. Data are clearly constrained in the
stable region (delimited by the solid line). These numerical results are in agreement with the solar wind analysis, as can be
seen comparing our study with Figure 2 of Ref. \cite{chen16}. This further confirms that the HVM model can describe the basic
dynamics of the solar wind ion species. However, in this direct comparison one has to take into account the limitations of the
approximation of isothermal (massless fluid) electrons, and limitations in the statistical convergence (data produced by three
numerical runs cannot fully recover the statistics of three years of spacecraft measurements). In panels (b) and (c) of Figure
\ref{fig2}, we show $\log_{ 10}\Lambda_f$ and $\log_{10}\Lambda_m$ respectively, as a function of $\log_{10}\beta$, $\beta$
being the total plasma beta. Here, one can easily see that the instability thresholds (black-dashed lines) are reached for
$\beta\geq 1$, again in good agreement with Figure 5 of Ref. \cite{chen16}.

\section{Coherent structures, kinetic effects and differential heating}
\begin{figure*}
\centering 
\includegraphics[width=7.5cm]{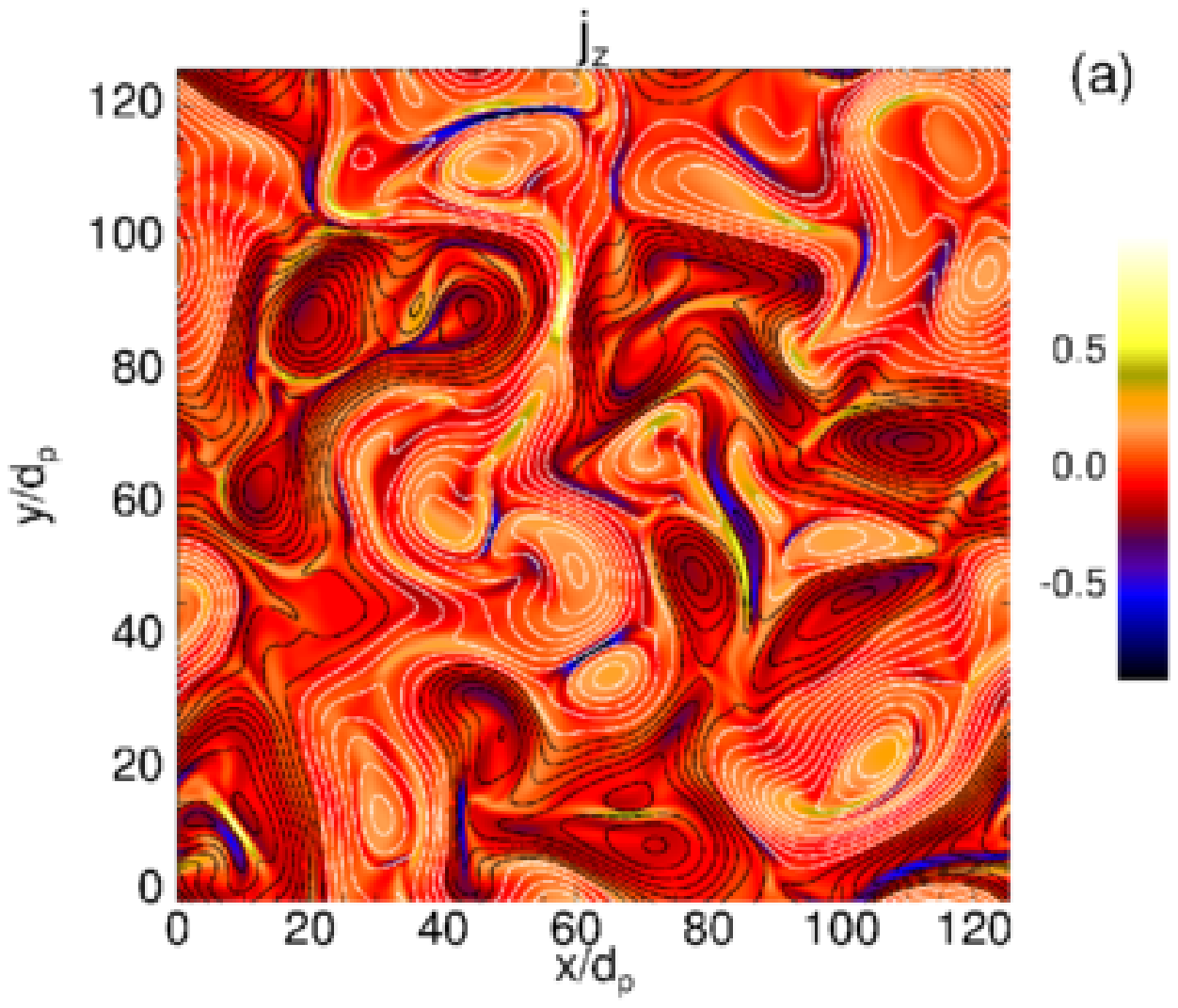}
\includegraphics[width=7.5cm]{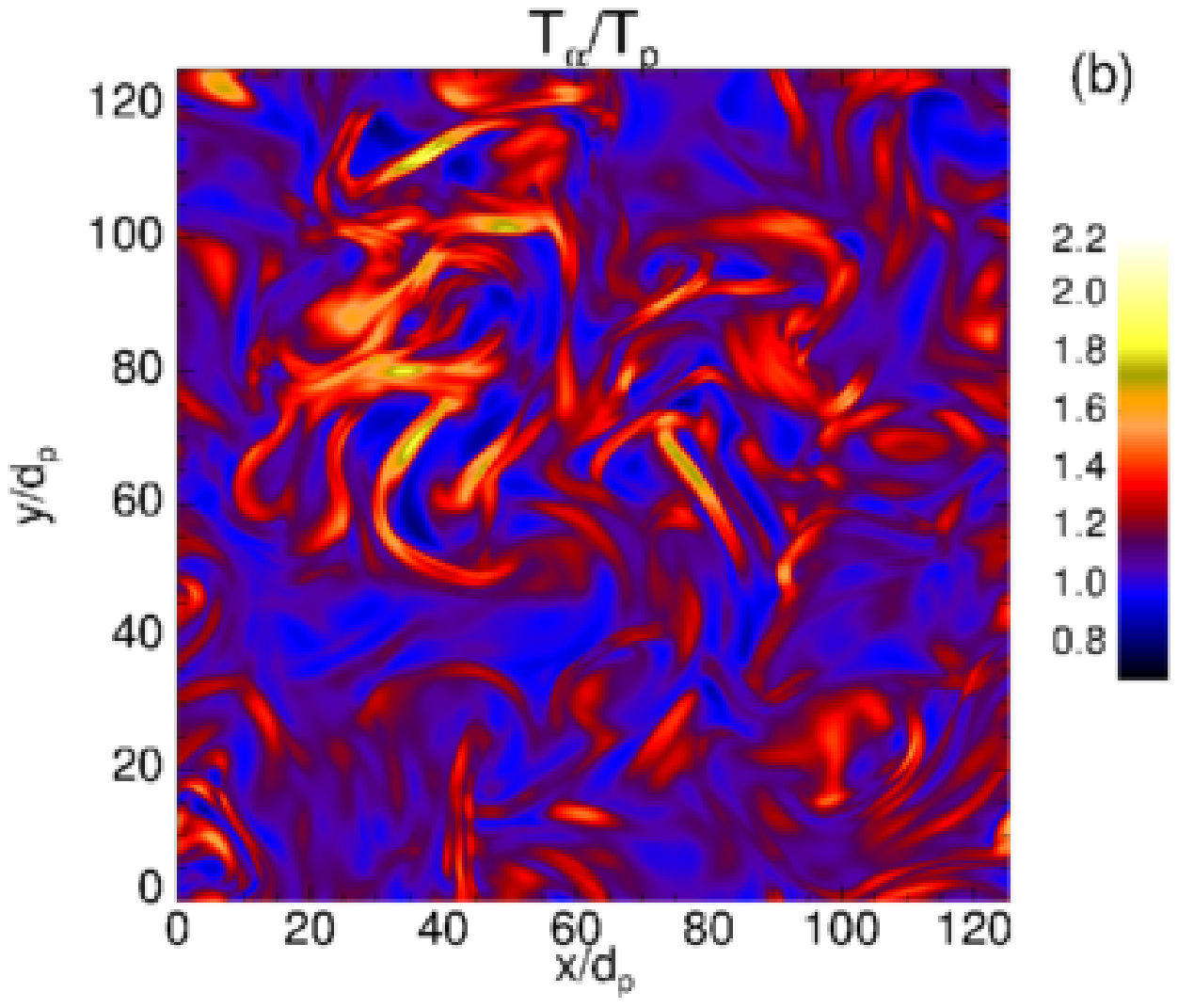}
\includegraphics[width=7.5cm]{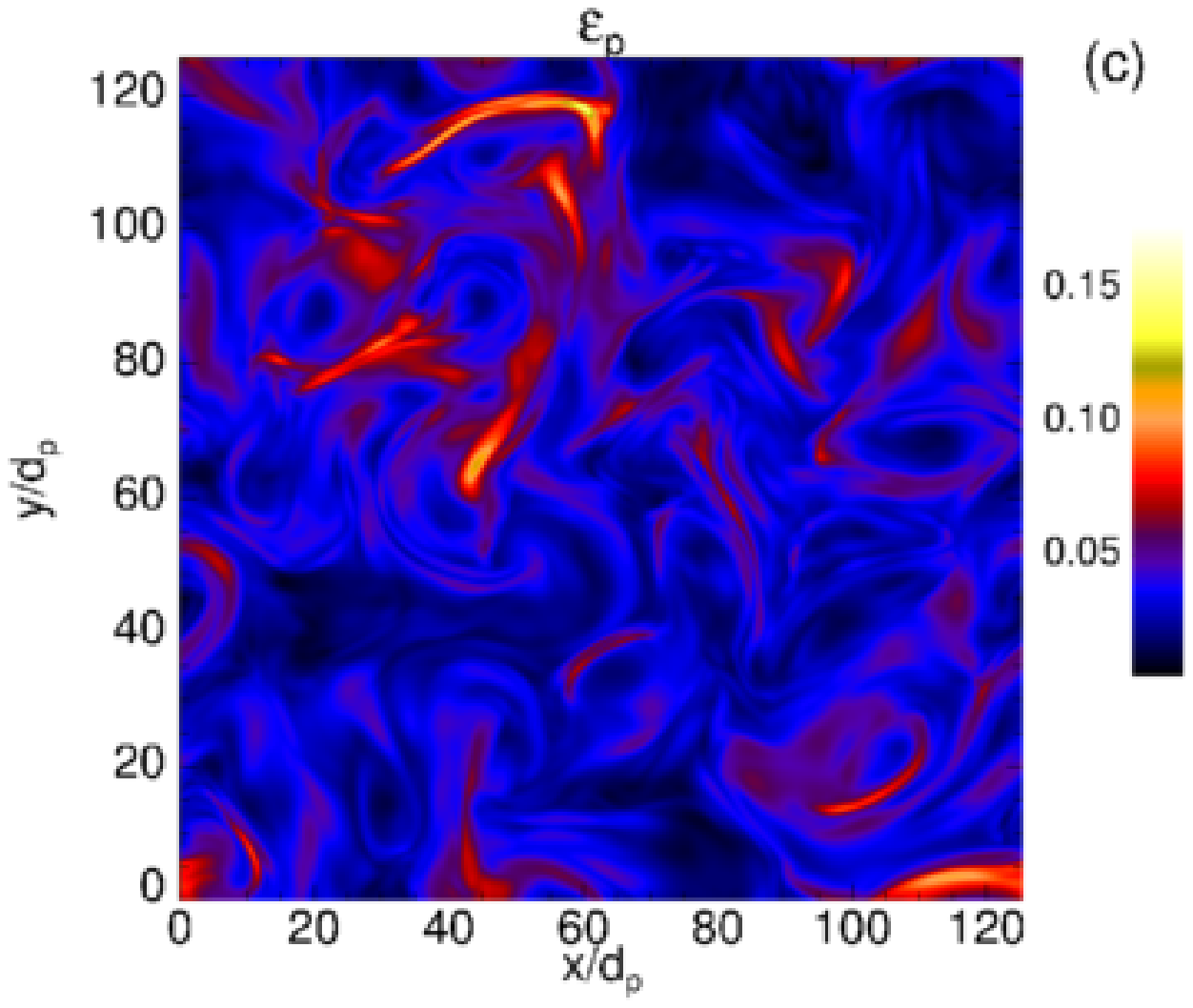}
\includegraphics[width=7.5cm]{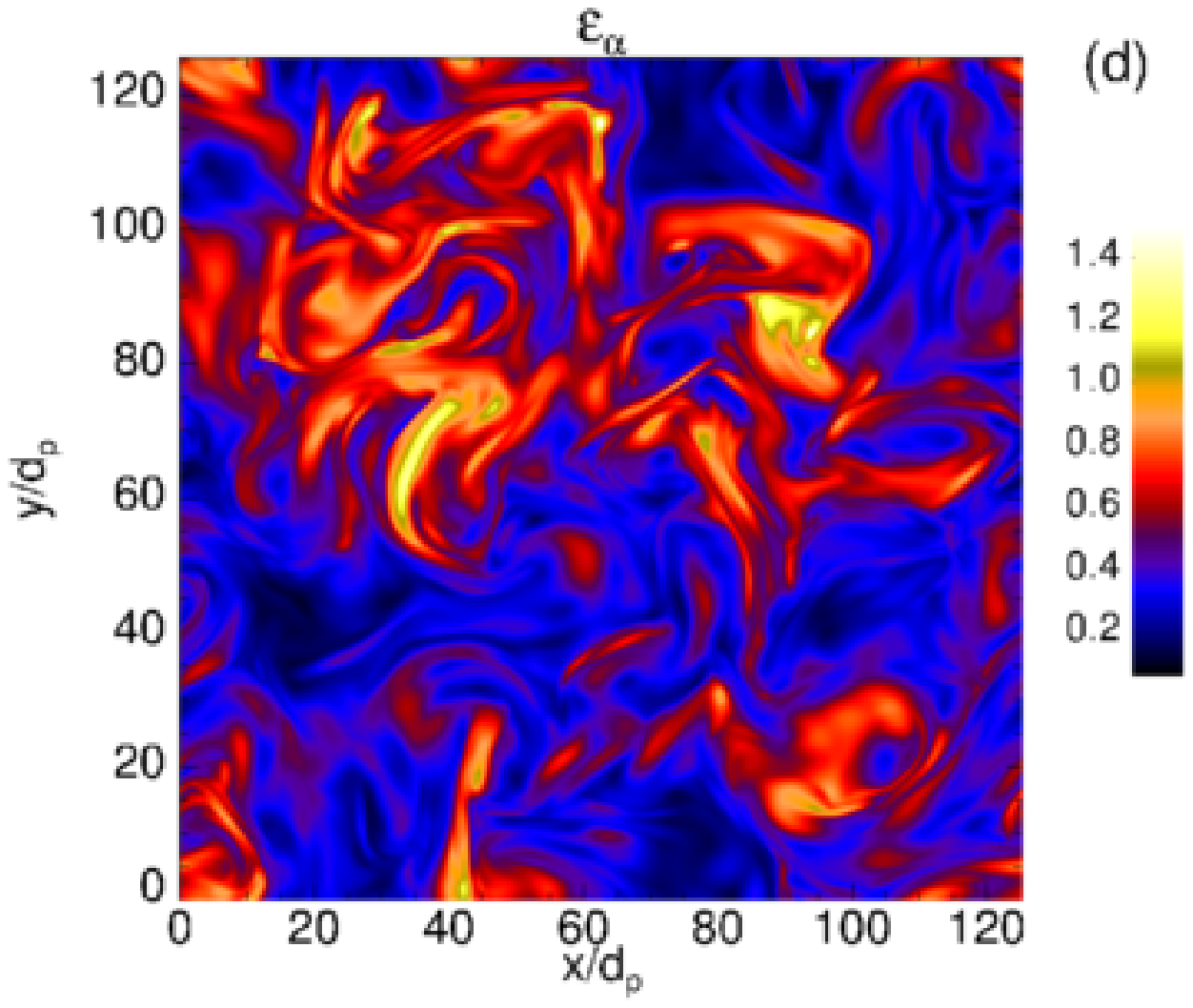}
\caption{(a) Out of plane current density $j_z$. The isolines of the magnetic potential $A_z$ are indicated by (black/white)
lines. (b) Ratio between alpha particle and proton temperatures. Panels (c) and (d): $\epsilon$ for both protons and alpha
particles, respectively. Each contour map is at the maximum of the turbulent activity.}
\lab{fig:contour}
\end{figure*}	

During the development of the turbulence cascade, temperature increase is observed for both ion species, alpha particles
being preferentially heated with respect to protons. Moreover, the generation of ion temperature anisotropy discussed in the
previous section is associated to the appearance of coherent structures, such as current sheets, filaments, strongly sheared
flows, etc., observed in the 2D patterns in physical space. In particular, panel (a) of Figure~\ref{fig:contour} displays the
shaded contours of the out of plane current density $j_z$ at $t=t^*$ for RUN 1, together with the contour lines of the magnetic
potential $A_z$ of the in-plane magnetic field $(\bB_{\perp} = \nabla A_z \times \be_z)$. Magnetic flux tubes with different
polarizations (or magnetic islands in 2D) are identified by closed contours of $A_z$, as it can be seen from panel (a) of this
figure. As it is clear from this contour plot, current density becomes very intense in between adjacent magnetic islands. We
point out that the mirror and fire hose unstable points in Figs. \ref{fig2} (b)-(c) (obtained in RUN 3, with increased
initial magnetic perturbations) are located in spatial positions very close to strong current sheets. In panel (b) of
Figure~\ref{fig:contour} we show the contour map of the proton to alpha particle temperature ratio. Even though at $t=0$ ions have
been initialized with equal temperature ($T_{0,p} = T_{0,\alpha}$), during the evolution of turbulence, the alpha particles
are heated preferentially with respect to protons. Moreover, the 2D map shows that the differential heating is not uniform, but is
strongly inhomogeneous, displaying a pattern similar to the 2D map of $j_z$ and being significantly influenced by the topology of
the magnetic field \cite{per14b,ser12,val14,ser15,wu13}. Moreover, the physical mechanism responsible for the ion heating process
is evidently more efficient for the alpha particles than for the protons.
\bfig
\centering 
\includegraphics[width=8.cm]{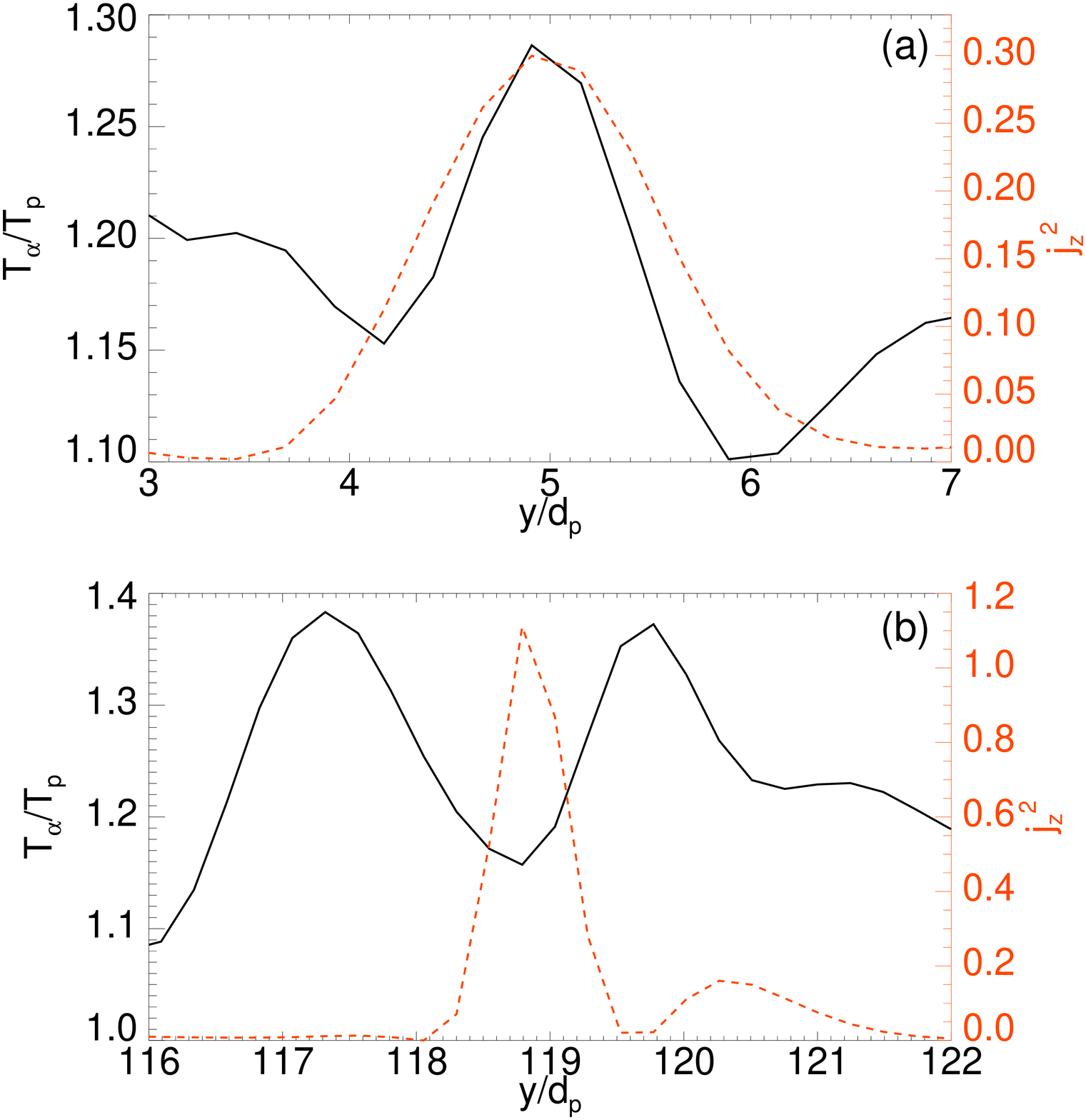}\includegraphics[width=8.cm]{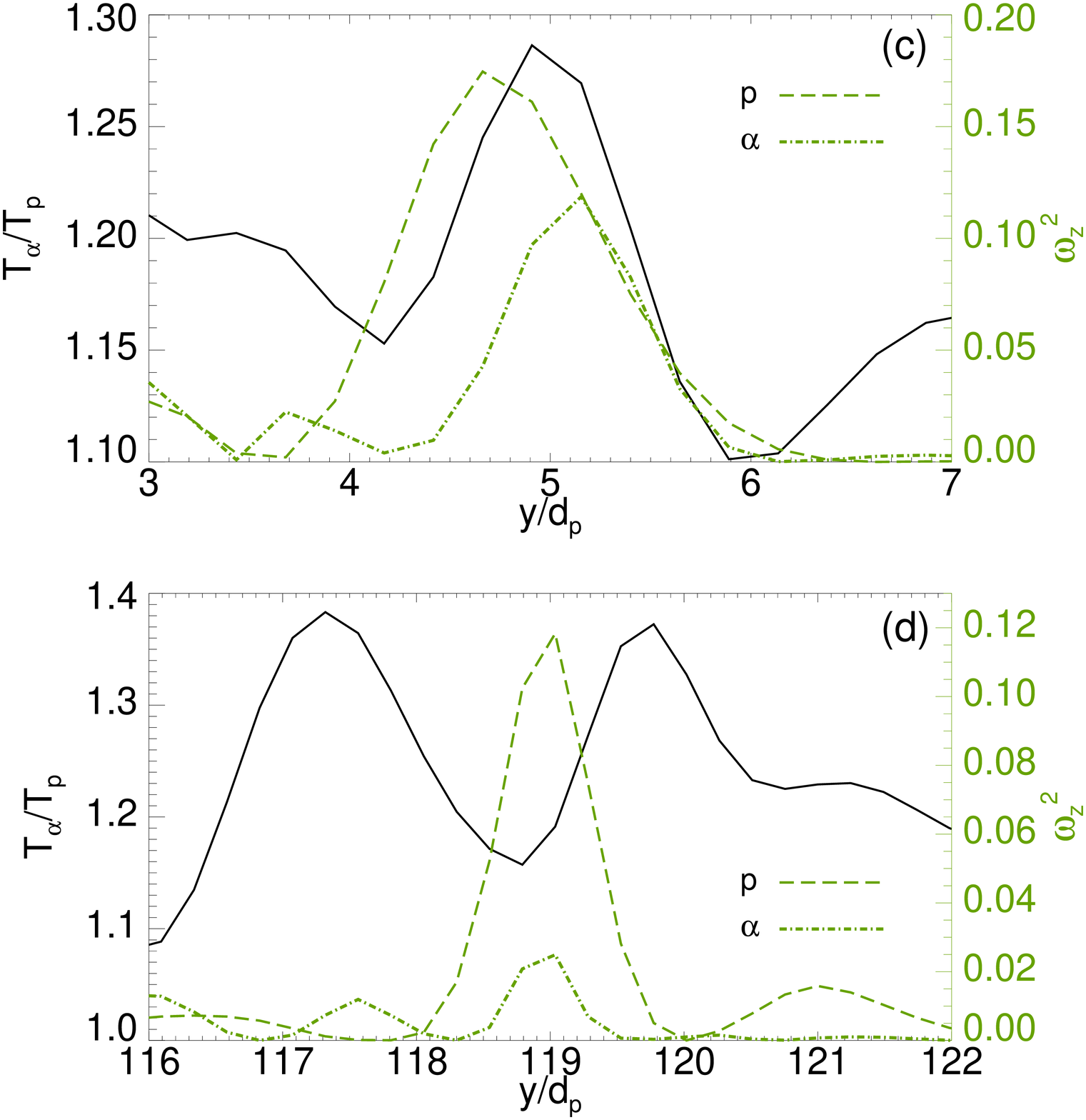}
\caption{Spatial profiles of the squared current density $j_z^2$ (red dashed line, panels (a) and (b)) and of the squared
vorticity $\omega_{z}^2$ for protons (green dashed line, panels (c) and (d)) and for alpha particles (green dot-dashed line,
panels (c) and (d)). In all panels, the alpha particle to proton temperature ratio is also shown (black solid line). All profiles
are evaluated along the $y$ direction at two different fixed spatial points: $x \sim 110 d_p$ (panels (a) and (c)) and $x \sim 50
d_p$ (panels (b) and (d)).}
\lab{fig:taglio}
\efig	

As the HVM code allows for a clean and almost noise-free description of the ion distribution function in phase space, we can
analyze the kinetic dynamics of the ion species associated to the development of the turbulent cascade and to the differential ion
heating. Several `in situ' observations  (see, for instance, Refs. \cite{mar06,mar12}) and previous numerical simulations
 \cite{val8,val9,per13,per14b,ser14,ser12,val14,ser15,per14a} have shown that the ion distribution functions in the turbulent
solar wind (especially in the less collisional fast wind) are typically far from thermodynamical equilibrium and that kinetic
effects manifest through the appearance of field-aligned beams of accelerated particles, generation of ring-like modulations in
the particle velocity distributions and, in general, through complicated non-Maxwellian deformations. In order to quantify the
deviation of the ion velocity distributions from the Maxwellian shape in our simulations, we introduce the following
non-Maxwellian measure \cite{gre12}, for each ion species: 
\beq
\label{eq:eps}
\epsilon_s = \frac{1}{n_s}  \sqrt{\int (f_s-g_s)^2 d^3 v} \ ,
\eeq
where $g_s$ is the associated equivalent Maxwellian distribution computed from the parameters of $f_s$ \cite{gre12}. We point out
that at $t=0$, $\epsilon_s$ is null, since both ion species have been initialized with a Maxwellian velocity distribution. We
observe that, simultaneously with the evolution of turbulence, the ion distributions dramatically deviate from the Maxwellian
shape. Panels (c) and (d) of Figure~\ref{fig:contour} display the contour maps of $\epsilon_s$ for protons and alpha particles,
respectively, at $t=t^*$ for RUN1. As it is clear from this figure, both ion species significantly deviate from thermodynamic
equilibrium and the spatial distribution of $\epsilon_s$ is not uniform, displaying features quite similar to the 2D contours of
the out of plane current density [panel (a)] and of the ion temperature ratio [panel (b)]. In particular, kinetic effects seem to
be localized near the peaks of the current density in the shape of thin filaments. It is worth noting that deviations from
Maxwellian are significantly stronger for the alpha particles than for the protons; that is, larger values of $\epsilon_\alpha$
are recovered with respect to $\epsilon_p$ and are achieved in spatial positions close to peaks of the current density, where
differential heating is also observed.
\begin{figure*}
\includegraphics[width=17cm]{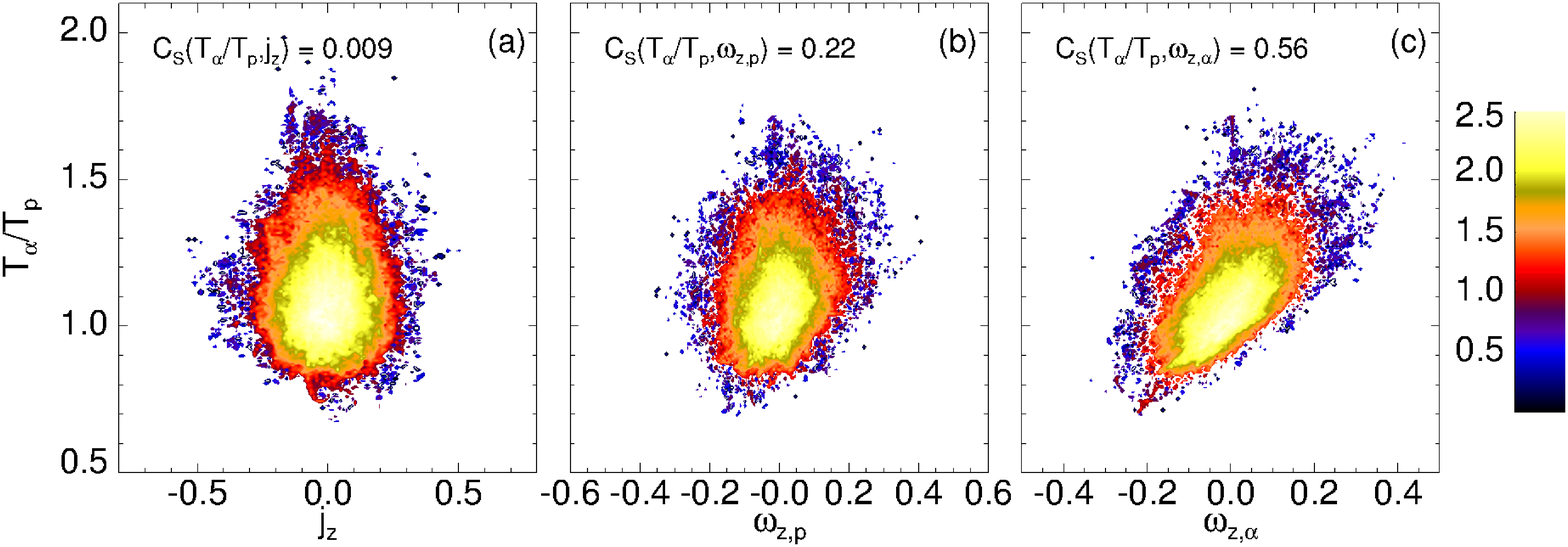}
\caption{From left to right: the joint probability distributions $P(T_\alpha/T_p,j_z)$, $P(T_\alpha/T_p,w_{z,p})$, and
$P(T_\alpha/T_p,\omega_{z,\alpha})$, in $\log_{10}$ scale. The corresponding Spearman correlation coefficients are indicated in
each plot.}
\lab{fig:correlation}
\end{figure*}

Ion differential heating usually occurs close to thin current sheets. Panels (a) and (b) of Figure \ref{fig:taglio} report 
the $y$ profile of $j_z^2$ (red-dashed curves), in two different ranges $y \in [3,7]d_p$ and $y\in[116,122]d_p$, computed at 
$x\sim 50 d_p$ and $x\sim 110 d_p$, respectively. Two current sheets of typical width of a few proton skin depths are visible
here. The black-solid curves in the same panels represent the corresponding $y$ profile of $T_\alpha/T_p$. As can be seen in the
two panels, differential heating is observed close to each strong current sheet. However, the behavior of the alpha to proton 
temperature ratio looks different in the two cases: in panel (a), alpha particles are preferentially heated in the center of 
the current sheet, while in panel (b) the differential heating occurs on both edges of the current sheet. This is a quite general 
behavior, in the sense that, close to each coherent current structure identified in the spatial domain, the alpha to proton 
temperature ratio displays a peak in the middle of the current sheet (as in panel (a)) or two peaks at its boundaries (as in 
panel (b)). Panels (c) and (d) of the same figure show similar plots for the squared vorticity component $z$ associated to the 
proton velocity, $\omega_{z,p}^2$, and to the alpha particle velocity, $\omega_{z,\alpha}^2$ where ${\bf\omega}=\nabla\times
{\bf v}$, at the same two positions as for the current $j_z^2$. The correlation observed for the vertical current is also
clearly present for the vorticity.
\begin{figure*}
  \centering 
\includegraphics[width=15cm]{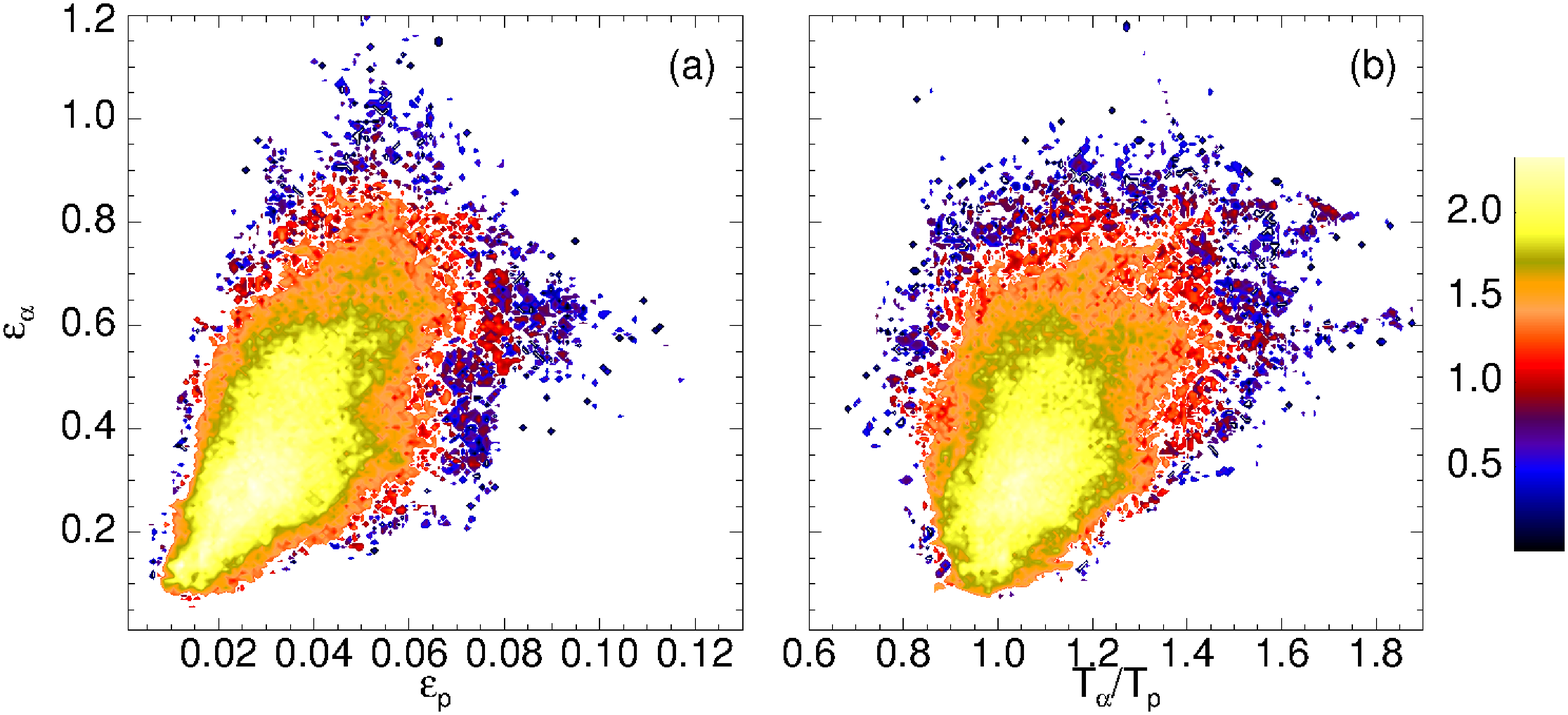}
\caption{Joint probability distributions  of $\epsilon_\alpha$ and $\epsilon_p$ (a) and of $\epsilon_\alpha$ and $T_\alpha/T_p$
(b), in $\log_{10}$ scale, for RUN 1 at $t=t^*$.}
\lab{fig:relation}
\end{figure*}	

The robustness of the correlation observed between the quantities shown in Figure~\ref{fig:taglio} can be visualized by means
of the joint probability distributions $P(T_\alpha/T_p,j_z)$ and similarly for the vorticity $P(T_\alpha/T_p,\omega_z)$, as shown
in Figure~\ref{fig:correlation}. In the left panel, it is evident that there is actually no statistical correlation between the
current and the differential heating, because of the spatial shift discussed above. On the contrary, an higher correlation is
visible between the proton vorticity and the differential heating (central panel of Figure~\ref{fig:correlation}). Even a stronger
correlation is observed with the alpha-particle vorticity (right panel). A more quantitative assessment can be obtained by
estimating the Spearman correlation coefficients between the alpha-proton temperature ratio and the current,
$C_S(T_\alpha/T_p,j_z)$, and similarly for the two vorticity fields, as reported in each panel of Figure~\ref{fig:correlation}.
This observation suggests that the vortical motion of the alpha particles may play a significant role in, or may be
significantly affected by, the differential heating process~\cite{franci}. This correlation can be understood also in terms of
magnetic reconnection. Topologically, magnetic reconnection might form quadrupolar vortical structures located near the current
peak, as clearly shown in \cite{Mat82}. In these regions, both magnetic and velocity shears can trigger non-Maxwellian and thermal
processes. A deeper analysis of the relationship of current and vortex structures with collisionless plasma heating can be
found in Ref. \cite{parashar16}.
\bfig
\centering 
\includegraphics[width=11cm]{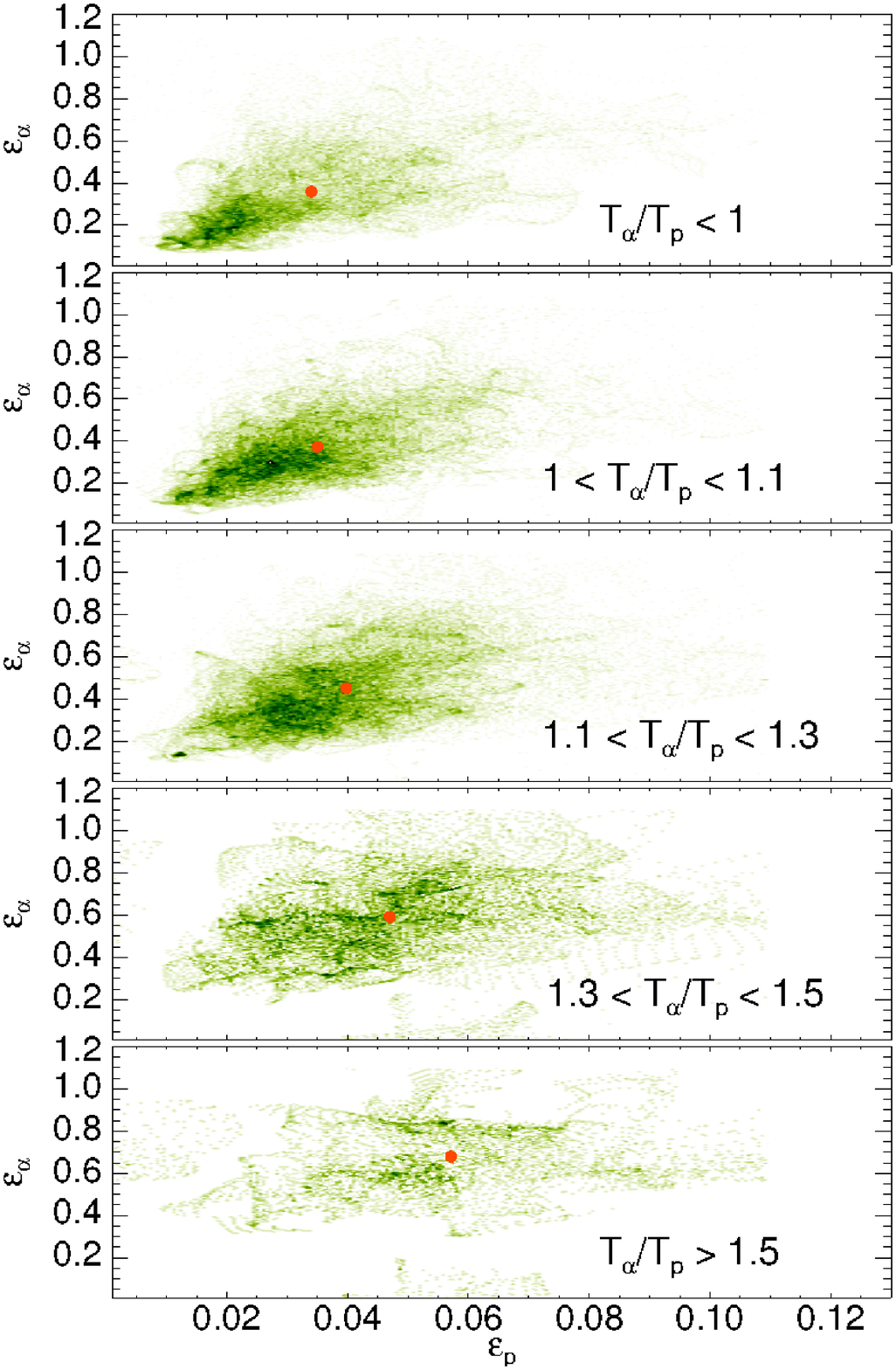}
\caption{Conditioned joint distributions $P(\epsilon_\alpha,\epsilon_p|T_\alpha/T_p)$, for five bins of the conditioning
variable $T_\alpha/T_p$ (increasing values from top to bottom); the location of the average value of $T_\alpha/T_p$ is
indicated by a red dot in each panel. Each distribution is normalized to its peak.}
\lab{fig:conditioned}
\efig	

At this point, it is important to better quantify the connection between the differential particle heating and the level of
kinetic activity. In Figure~\ref{fig:relation} we show the joint distributions of $\epsilon_\alpha$ and $\epsilon_p$ (left panel)
and of $\epsilon_\alpha$ and $T_\alpha/T_p$ (right panel) for RUN 1 at $t=t^*$. Both panels show that (i) alpha and proton
velocity distributions tend to deviate from Maxwellian at similar spatial locations, and (ii) such a deviation is correlated
with the differential heating. In Figure \ref{fig:conditioned}, the joint distribution of $\epsilon_p$ and $\epsilon_\alpha$
has been conditioned to specific ranges of values of the temperature ratio $T_\alpha/T_p$, divided into five bins as indicated in
each panel. Each distribution is normalized to its peak value. As the temperature ratio increases (from top to bottom in the
figure), an evident shift of the distribution towards larger values of both $\epsilon_p$ and $\epsilon_\alpha$ is observed; this
shift can be clearly appreciated by looking at the location of the average value of $T_\alpha/T_p$, indicated by a red dot in each
panel of this figure. These results indicate that a larger deviation from a Maxwellian distribution is observed for larger
temperature ratios $T_\alpha/T_p$. On the other hand, the conditioned joint distributions are relatively symmetric, so that there
is no evidence of different increase in $\epsilon_p$ or $\epsilon_\alpha$ when the alpha particles are preferentially heated.

Temperature variations of the ions seem to be related to their kinetic dynamics and, therefore, to the complex deformations
of the particle velocity distributions. Future space missions, such as THOR \cite{vaivads16}, designed to provide measurements of
the three-dimensional ion distribution functions with very high phase space resolution, will be able to capture the non-Maxwellian
deviations of the velocity distributions, so as to provide relevant clues on the long-standing problem of particle heating.

As discussed previously, close to the sites of enhanced magnetic activity, the non-Maxwellian measure $\epsilon_s$ achieves large
values for both ion species. In order to show what the three-dimensional ion velocity distributions look like in turbulence, in
Figure~\ref{fig:vdf} we report the iso-surfaces of the proton [panel (a)] and alpha particle [panel (b)] velocity distributions,
computed in the spatial point where $\epsilon_s$ is maximum. The red tube in both panels indicates the direction of the local
magnetic field. The proton velocity distribution displays a field-aligned beam and the generation of ring-like modulations in
planes perpendicular to the direction of the local field, presumably signatures of cyclotron resonance. On the other hand, for
the alpha particles we recovered more evident deformations, with the appearance of multiple blobs and elongations; the alpha
particle velocity distribution seems to lose any property of symmetry with respect to the direction of the local magnetic field.
It is clear from these plots that kinetic effects, working along the turbulent cascade at typical ion scales, can efficiently
drive the plasma away from thermodynamic equilibrium. Moreover, when the particle velocity distributions are not constrained
by limiting assumptions on their shape in velocity space, they can be distorted in a very complicated way and lose their
symmetry.

\section{Deviations from Maxwellian distributions}
An alternative way to describe the deviations from thermodynamic equilibrium consists in defining the so-called minimum variance
frame of the three-dimensional velocity distribution. We compute the preferred directions of the velocity distribution of each
ion species in velocity space \cite{ser12}, in each spatial position, from the stress tensor:
\begin{equation}
 \Pi_{i j}=n^{-1} \int (v_i - u_i)(v_j - u_j)f d^3 v,
 \label{pressure}
\end{equation}
\begin{figure*}
  \centering 
\includegraphics[width=14cm]{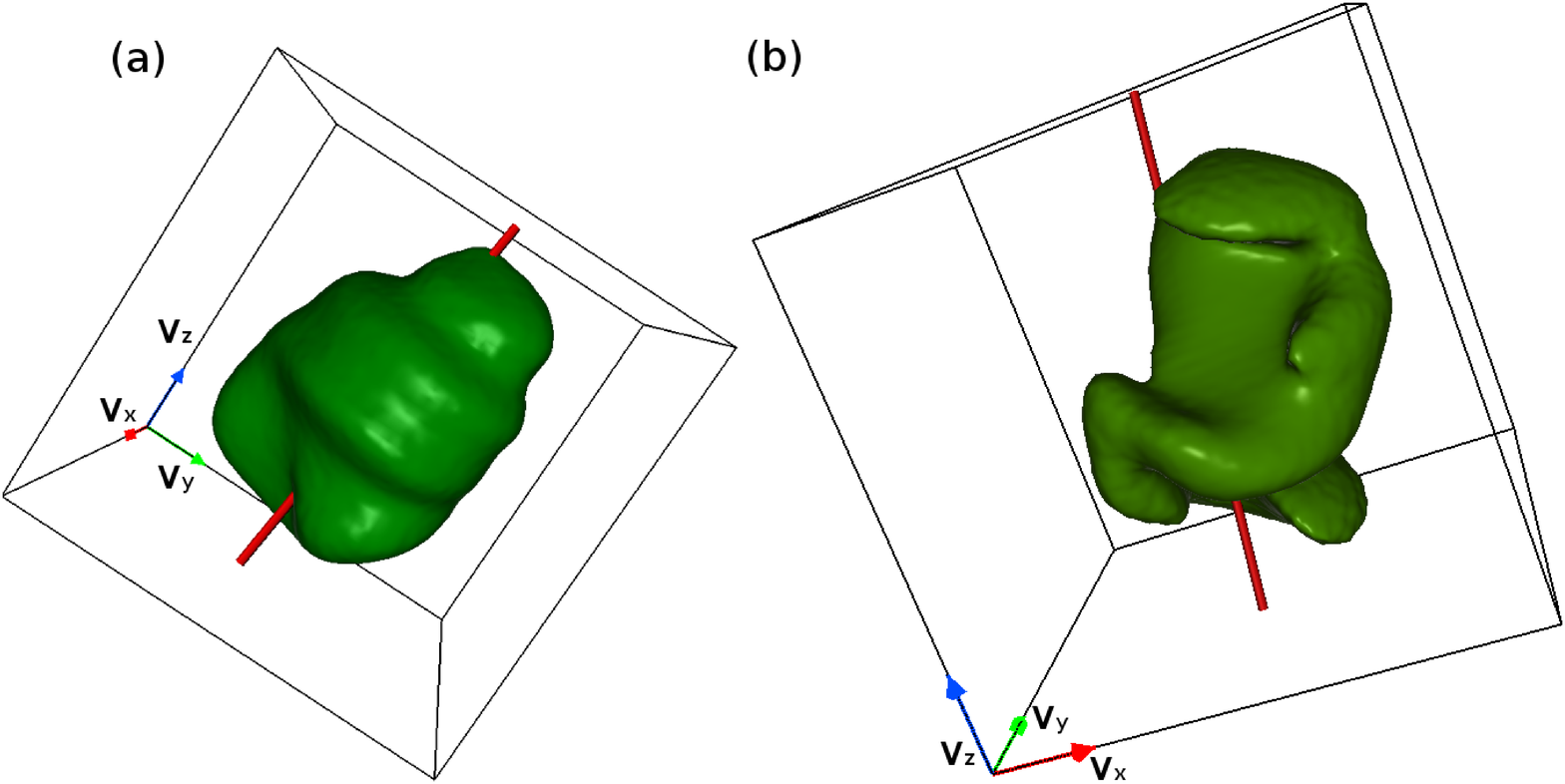}
\caption{Iso-surfaces of the proton (panel (a)) and alpha particle (panel (b)) velocity distribution in a spatial point where
$\epsilon_s$ is maximum. The direction of the local magnetic field vector is reported as a red tube.}
\lab{fig:vdf}
\end{figure*}	

This tensor can be diagonalized by computing its eigenvalues $\{\lambda_1, \lambda_2, \lambda_3\}$ (ordered in such way that
$\lambda_1 >\lambda_2 > \lambda_3$) and the corresponding normalized eigenvectors $\{ {\bf \hat{e}}_1, {\bf \hat{e}}_2,  {\bf
\hat{e}}_3\}$. We point out that $\lambda_i$ are proportional to temperatures and ${\bf \hat{e}}_i$ are the anisotropy
directions of the velocity distribution. 

The quantity $\epsilon_s$ gives general information on the departure of the ion species from thermodynamic equilibrium, while
the eigenvalues $\lambda_i$ provide insights into the properties of symmetry of the ion velocity distributions. Therefore, for RUN
1 at $t=t^*$, we computed the PDF of the ratios $\lambda_i/\lambda_j$ ($i,j=1,2,3$ and $j\ne i$), conditioned to the values of
$\epsilon_s (t=t^\ast)$. Note that each of these ratios is equal to unity for a Maxwellian velocity distribution. In Figure
\ref{fig:lambda}, we show the PDF of $\lambda_1/\lambda_2$ (left panels), $\lambda_1/\lambda_3$ (middle panels) and
$\lambda_2/\lambda_3$ (right panels) for protons (top row) and alpha particles (bottom row); these PDFs have been computed for
three different ranges of values of $\epsilon_s$, $0\leq\epsilon_s\leq\hat{\epsilon}_{1,s}$ (black curve),
$\hat{\epsilon}_{1,s}<\epsilon_s\leq \hat{\epsilon}_{2,s}$ (red curve) and $\hat{\epsilon}_{2,s}<\epsilon_s\leq\epsilon_{s,max}$
(blue curve), where $\epsilon_{s,max}$ represents the maximum value of $\epsilon_s$ over the two-dimensional spatial domain at
$t=t^*$ ($\epsilon_{p,max}=0.17$, $\epsilon_{\alpha,max}=1.5$) and $\hat{\epsilon}_{1,p}=0.04$, $\hat{\epsilon}_{2,p}=0.07$,
$\hat{\epsilon}_{1,\alpha}=0.5$, $\hat{\epsilon}_{2,\alpha}=1.0$.
\begin{figure*}
  \centering 
\includegraphics[width=15cm]{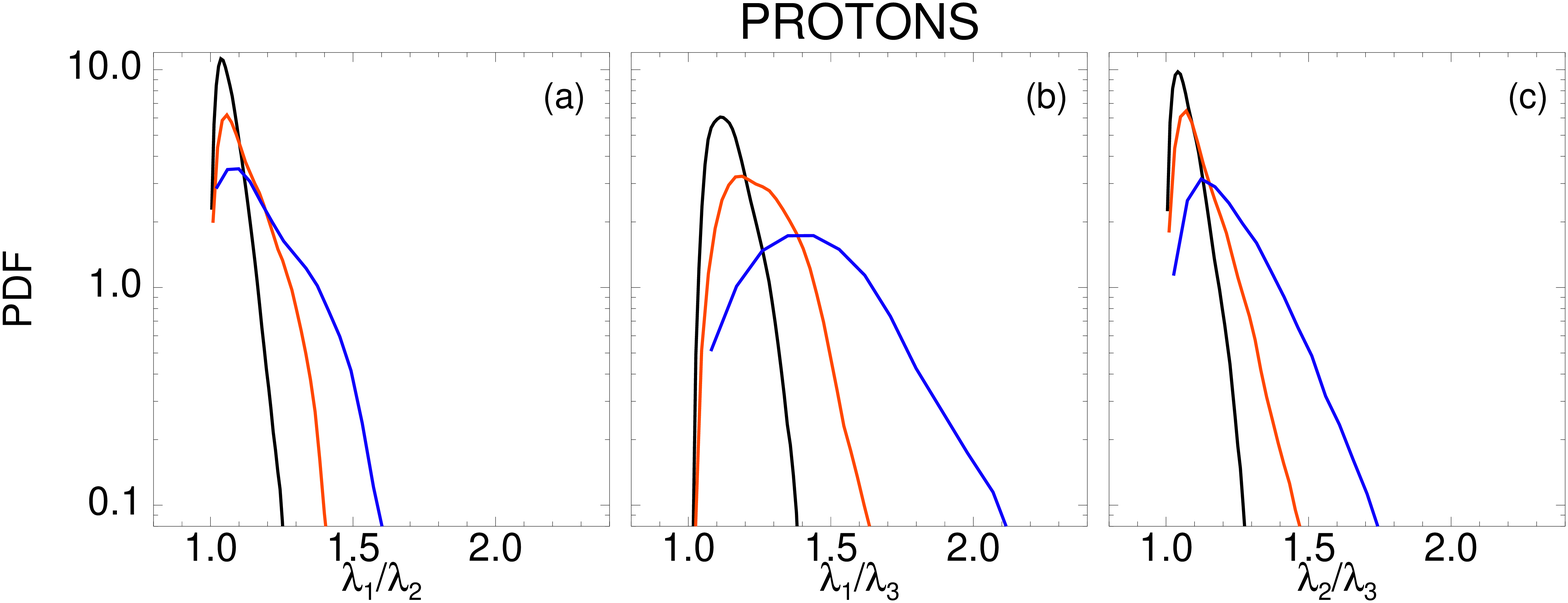}
\includegraphics[width=15cm]{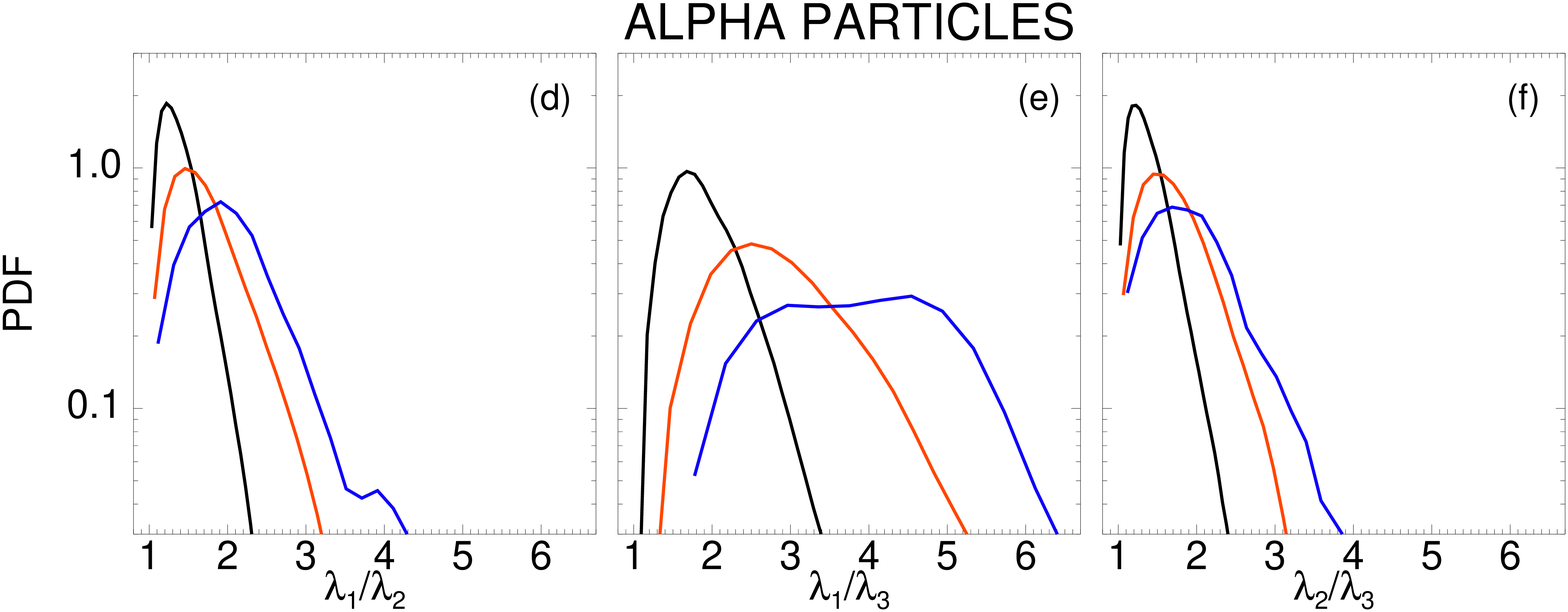}
\caption{PDF of $\lambda_1/\lambda_2$ (panels (a) and (d)), $\lambda_1/\lambda_3$ (panels (b) and (e)) and $\lambda_2/\lambda_3$
(panels (c) and (f)) for both protons (top panels) and alpha particles (bottom panels). Each PDF has been computed for three
different ranges of values of $\epsilon_s$, namely $0\leq\epsilon_s\leq\hat{\epsilon}_{1,s}$ (black
curve), $\hat{\epsilon}_{1,s}<\epsilon_s\leq \hat{\epsilon}_{2,s}$ (red curve) and
$\hat{\epsilon}_{2,s}<\epsilon_s\leq\epsilon_{s,max}$ (blue curve).}
\lab{fig:lambda}
\end{figure*}	

From this latter figure it can be noticed that, in the range of small $\epsilon_s$, the PDFs (for both protons and alphas) have a
peak close to unity (they are not exactly centered around $1$, since the minimum value of $\epsilon_s$ is not zero at $t=t^*$). 
As $\epsilon_s$ increases, high tails appear in the PDFs suggesting that, in the case of significant deviations from
Maxwellian, the ion velocity distribution loses its properties of isotropy and gyrotropy. This observation suggests that the use
of reduced models, based on restrictive approximations on the shape of the ion velocity distribution, might not be appropriate,
and more complete models, describing the evolution of the velocity distributions in a full 3D velocity space, should be employed.

With the aim of characterizing the nature of the deformation of the particle velocity distributions and to identify the spatial
regions which are the sites of kinetic activity, we computed for each ion species two indices of departure from Maxwellian, i. e.
the temperature anisotropy index and the gyrotropy index, in two different reference frames, namely the MVF and the local magnetic
field frame (LMF). We define the anisotropy measures $\zeta=|1-\lambda_1/\lambda_3|$ (MVF) and
$\zeta^\ast=|1-T_\perp/T_\parallel|$ (LMF), where $T_\perp$ and $T_\parallel$ are the temperatures with respect to the local
magnetic field, and the gyrotropy indicator in the MVF $\eta=|1-\lambda_2/\lambda_3|$. The gyrotropy index in the LMF $\eta^\ast$
can be computed by using the normalized Frobenius norm of the nongyrotropic part $\Nv$ of the full pressure tensor ${\bf\Pi}$
\cite{aun13}: 
\begin{equation}
 \eta^\ast = \frac{\sqrt{\sum_{i,j} N_{ij}^2}}{Tr({\bf\Pi})},
\end{equation}
\noindent
where $N_{ij}$ are the components of the tensor $\Nv$, and $Tr (\Nv) = 0$. It is worth to point out that all indices defined above
are identically zero if the particle velocity distribution is Maxwellian. At the maximum of the turbulent activity, these indices
significantly differ from zero for both ion species, meaning that the ion velocity distributions are anisotropic and
non-gyrotropic 
in both MVF and LMF.
\begin{figure*}
\includegraphics[width=8cm]{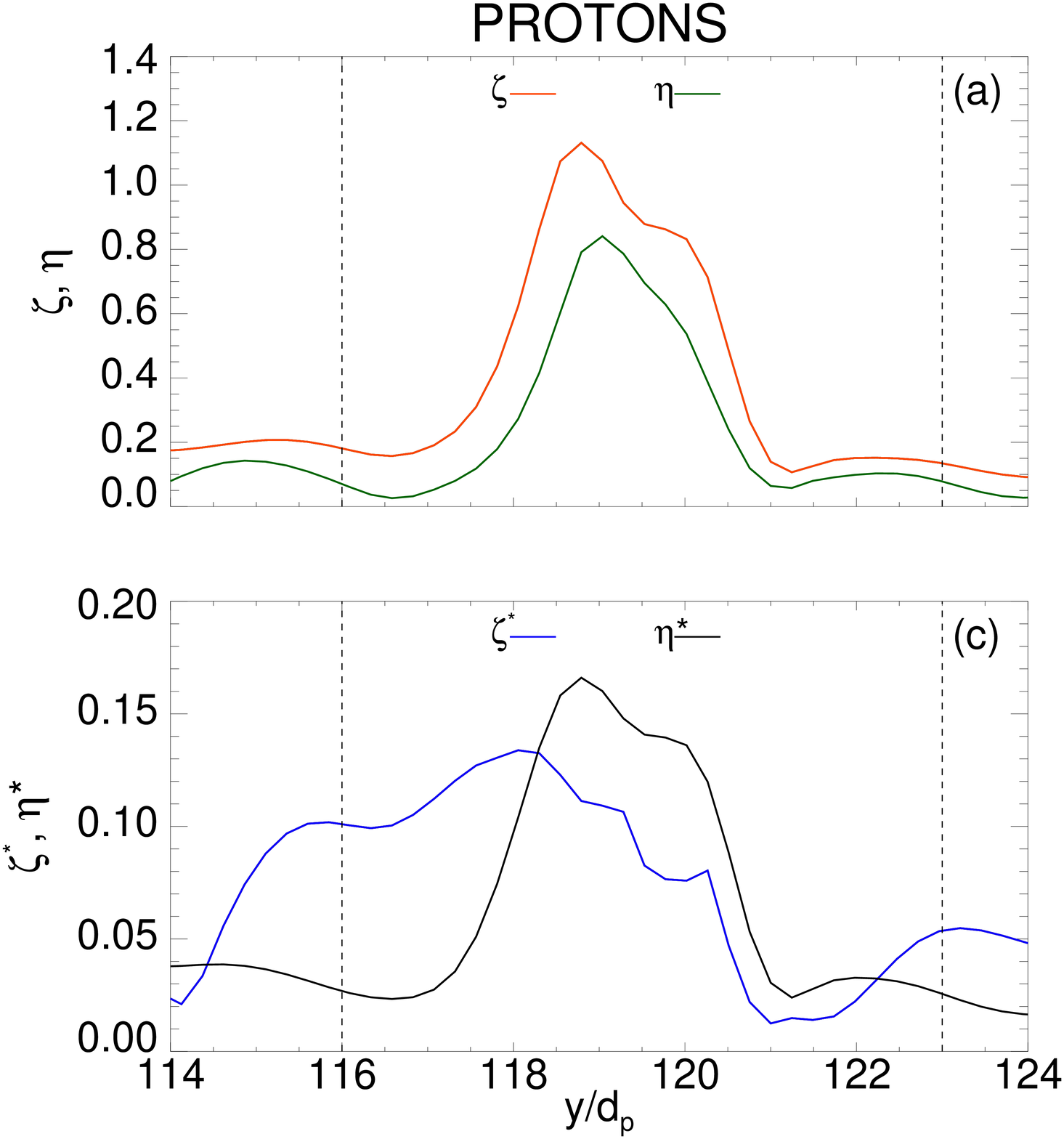}
\includegraphics[width=8cm]{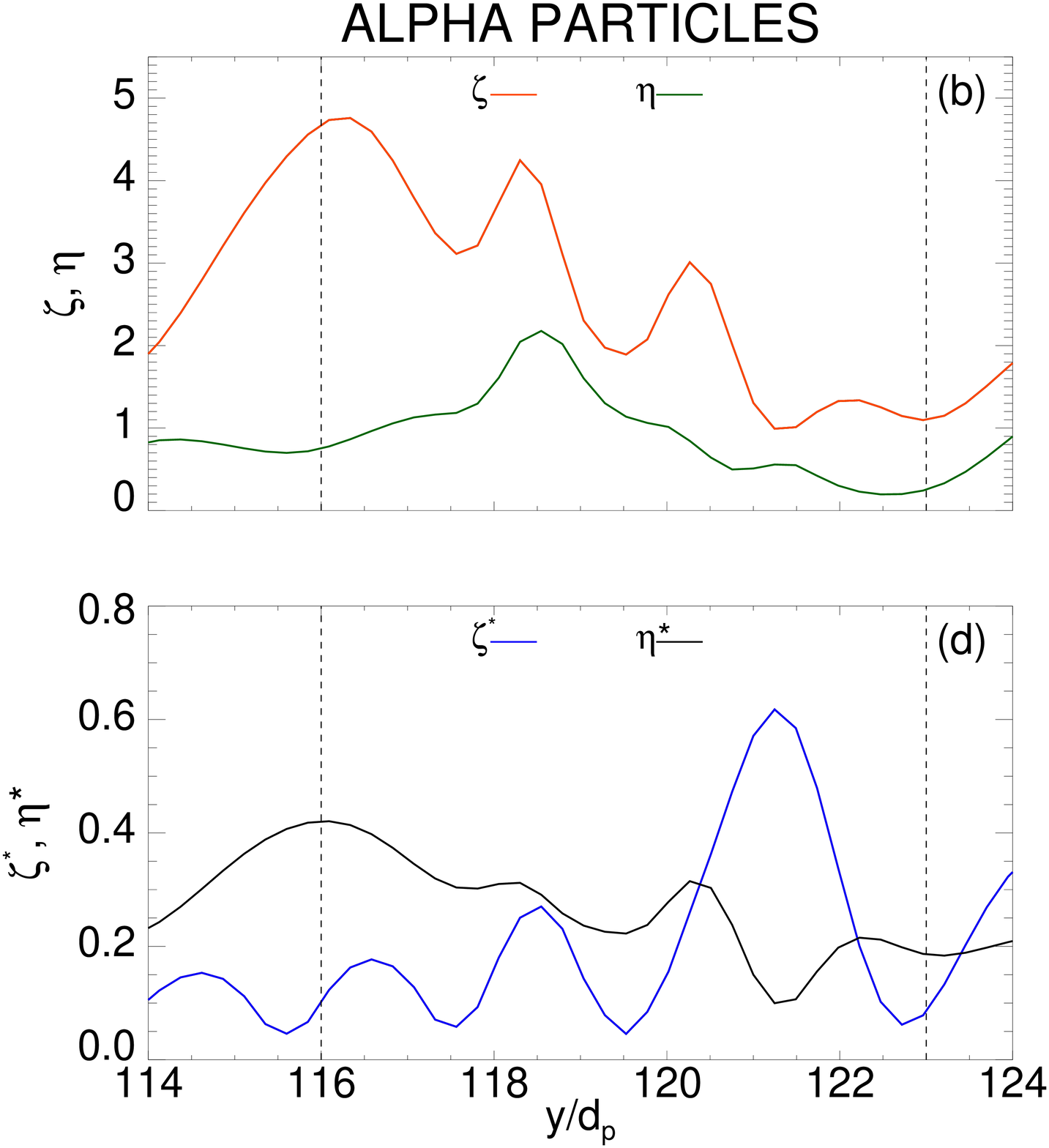}
\caption{Temperature anisotropy indices $\zeta$ (red line) and $\zeta^*$ (blue line) and non-gyrotropy indices $\eta$ (green
line) and $\eta^*$ (black line), for protons (left panels) and alpha particles (right panels), inside the current sheet in
panel (b) of Figure \ref{fig:taglio}. Top and bottom panels refer to MVF and LMF, respectively.}
\lab{fig:index_taglio}
\end{figure*}	

In Figure \ref{fig:index_taglio}, we show the spatial profiles of all these indices inside the region where the current sheet
shown in Figures~\ref{fig:taglio} (b) is located (indicated by vertical black-dashed lines). The indices for protons are reported
in the left panels, while for alpha particles in the right panels. As it is evident from this figure, the non-Maxwellian
measures of both protons and alpha particles are not null in the regions close to the peak of the current density, in
both reference frames; specifically, the ion velocity distributions become anisotropic and non-gyrotropic, as can be seen from
both the MVF and LMF indices. It is worth noting that the indices for the alpha particles are significantly larger than for the
protons, additional evidence that alpha particles are more efficiently driven out of equilibrium with respect to protons.

\section{Optimized measurements for the study of protons and alpha particles kinetic dynamics}
In the previous sections we have shown that kinetic effects produce highly non Maxwellian distributions, particle beams and
ring-like modulations. An instrument designed to study these effects should have proper phase space (and temporal) resolution,
since a poor resolution could impede the observations of such fine features. In near-Earth space it is possible to
perform high resolution in situ plasma observations. Measurements of particle velocity distribution functions are usually
performed by top-hat electrostatic analyzers \cite{Carlson}. Such instruments are composed of two concentric hemispheres set at
different voltage, with an aperture on the outer hemisphere.  The electric field between the plates allows for particles within a
specific energy-per-charge (E/q) ratio to pass through the gap and reach the detector. Varying the voltage permits to sort
particles according to their E/q. Parallel incident particles will be focused on a specific sector of the detector at the analyzer
exit, each sector identifying the particle velocity azimuthal direction. Analyzers of this type measure incoming particles with a
$360^\circ$ disk shaped field of view. In order to sample the entire $4\pi$ solid angle, either the spacecraft rotation is used or
electrostatic polar deflectors are employed. Usually analyzers for solar wind measurements are characterized by a restricted
field of view, since the solar wind at 1 AU is a cold beam in velocity space. A schematic illustration of such a top-hat
analyzer is given in
figure~\ref{fig:tophat}.
\begin{figure*}
\includegraphics[width=14cm]{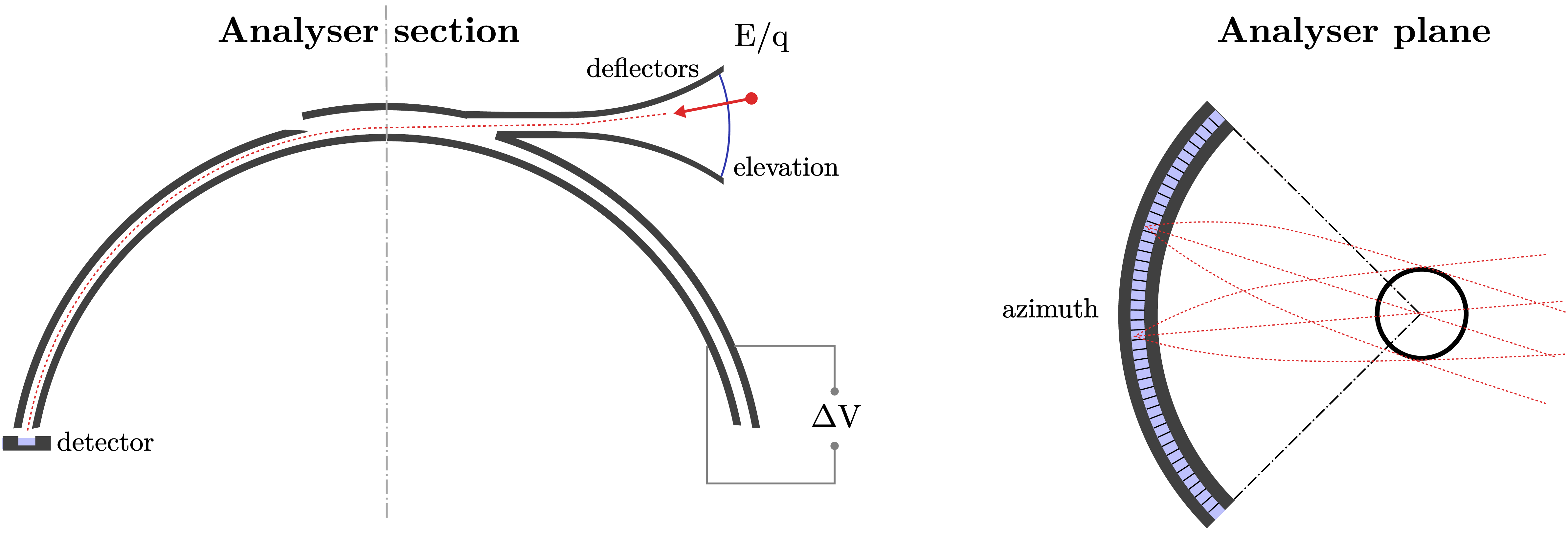}
\caption{Schematic representation of a top-hat electrostatic analyzer specifically designed for measuring the solar wind proton
and alpha distributions \cite{lavraud16}.}
\lab{fig:tophat}
\end{figure*}

As already said, the energy-angular resolution of the top-hat analyzer is extremely important when aiming to study the signatures
of kinetic-scale processes, since low phase space resolution would smooth out the fine structures of the particles distribution
functions. The Cold Solar Wind (CSW) \cite{lavraud16} on board the THOR mission is a top-hat analyzer conceived to measure solar
wind protons and alpha particles distribution functions with unprecedented temporal and phase space resolutions
\cite{demarco16,lavraud16}. In particular, it has an angular and energy resolution that will permit to observe the fine structure
characterizing the particle distribution functions related to the effects of turbulence.

In this respect, we use a \emph{top-hat simulator} to find out how the proton and alpha velocity distributions characterized by
large $\epsilon_s$, and shown in Figure \ref{fig:vdf}, would be detected by CSW. Such distribution functions are meaningful
examples since they are characterized by non Maxwellian shapes and a high degree of complexity as has been shown in the previous
sections. Moreover, we compare the moments and the $\epsilon_s$ of the HVM and CSW distributions in proximity of large
$\epsilon_s$ sites, showing that CSW measurements capture the complexity present in the HVM distributions.  

CSW will detect particles in 96 energy-per-charge intervals with $\Delta E/E=7\%$. It will have a restricted field of view of
$48^\circ$ in both azimuth and elevation, focused on the cold solar wind population, and will have an unprecedentedly high angular
resolutions of $1.5^\circ$, in both azimuth and elevation. The CSW response is further determined by its geometric factor
$G=2.2\times 10^{-5}\mbox{cm}^2\mbox{sr}$ and the time needed for the accumulation of particle counts in one energy channel and
elevation angle $T_{acc} \sim 0.25\mbox{ms}$.
\begin{figure*}
\includegraphics[width=8cm]{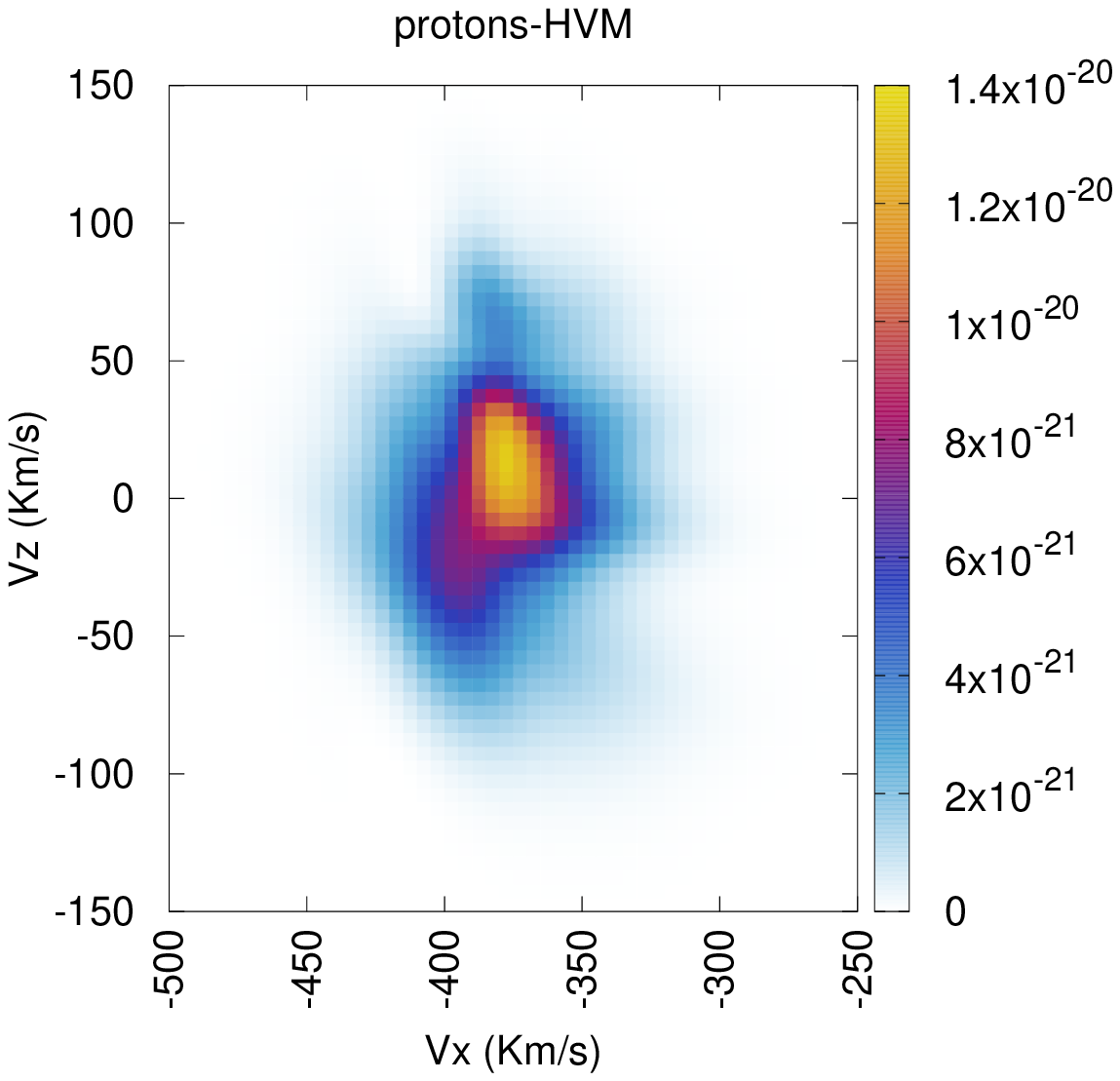}
\includegraphics[width=8cm]{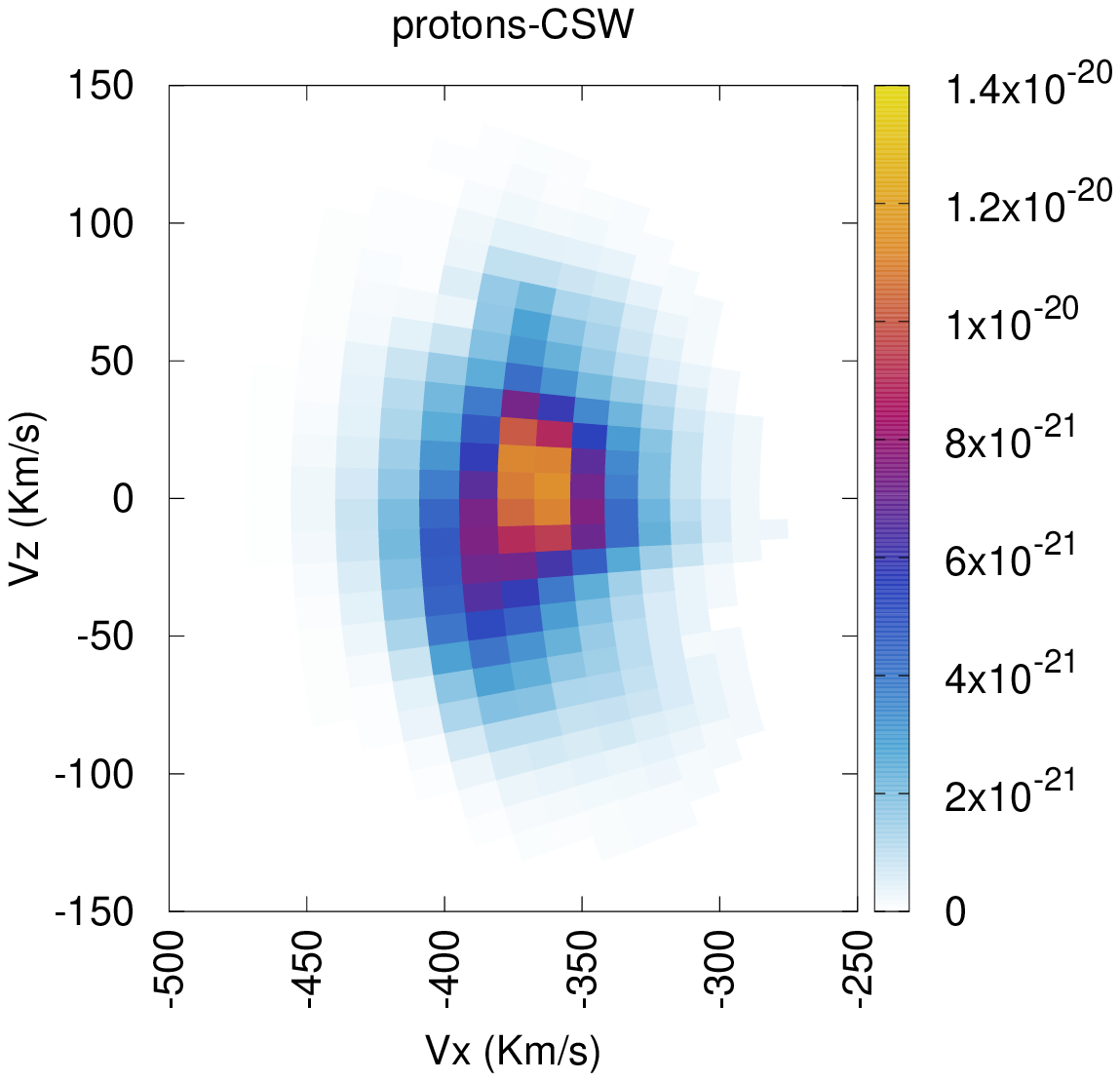}
\includegraphics[width=8cm]{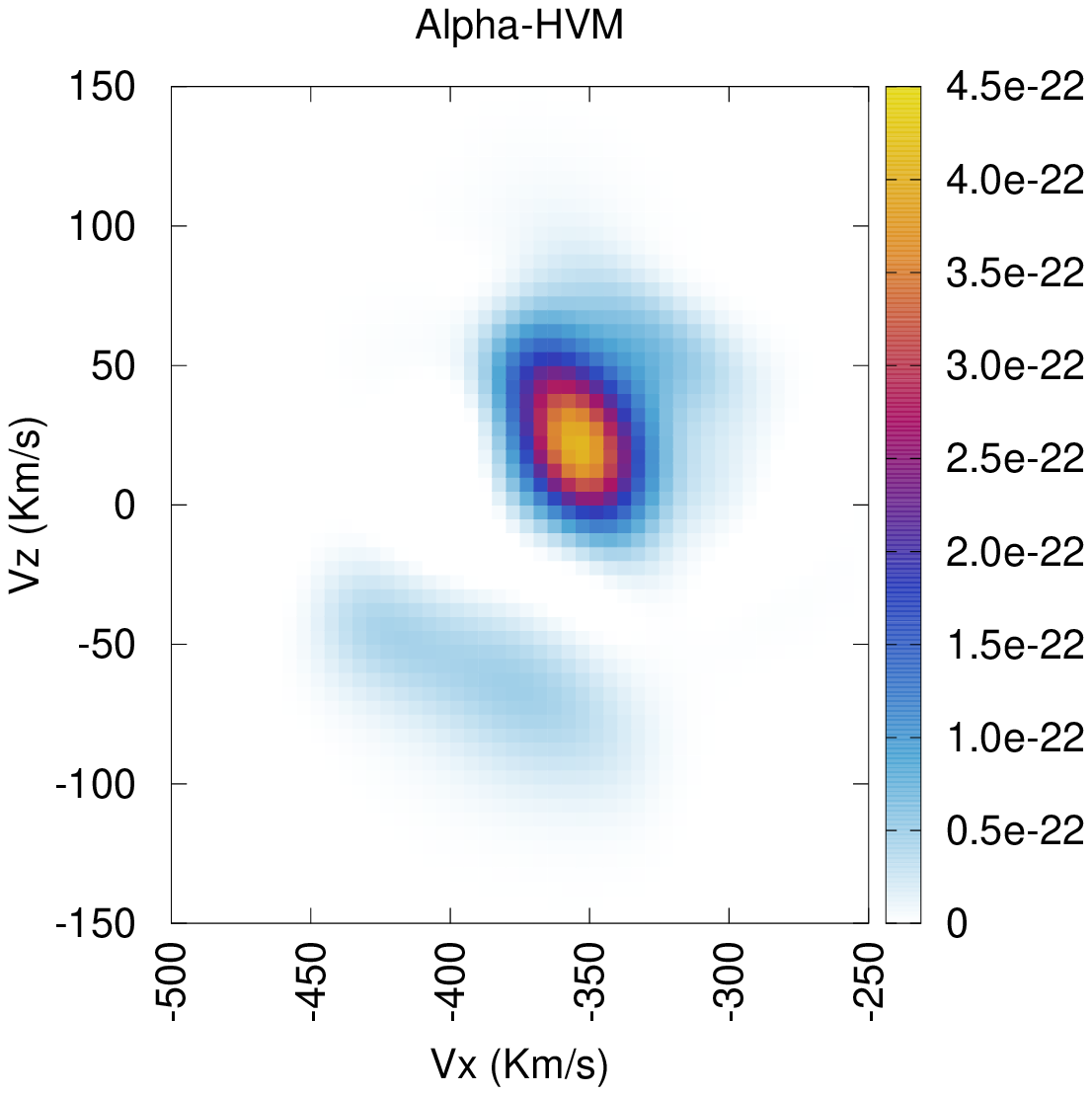}
\includegraphics[width=8cm]{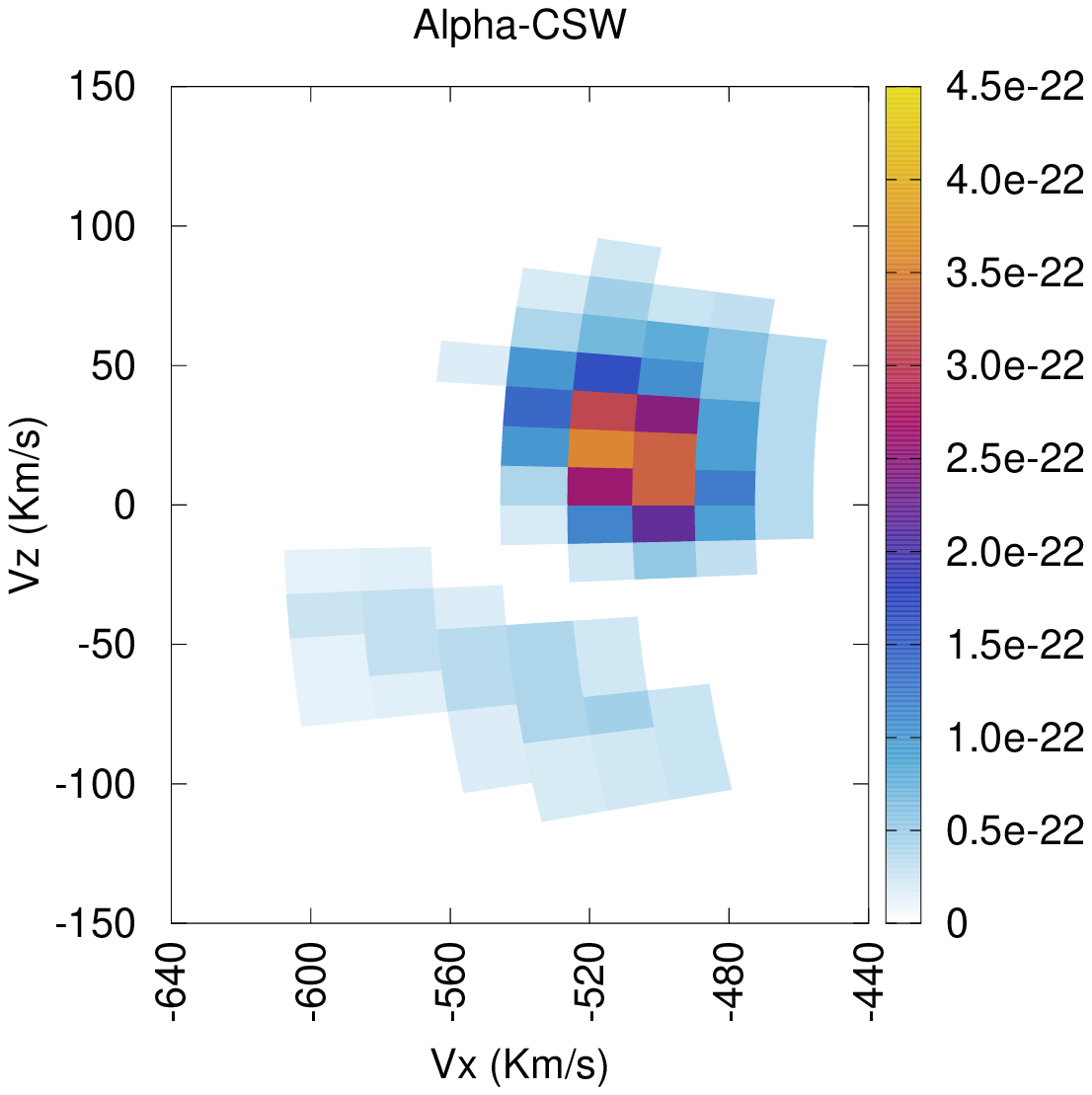}
\caption{Protons and alpha particles velocity distributions as recovered from the simulations and as sampled by CSW. Note that the
distribution function scale is not the same for protons and alphas.}
\lab{fig:tophatsim}
\end{figure*}

As a first step, the HVM distributions are expressed in physical units by rescaling the normalized  variables of the simulation to
physical quantities. In order to adapt to a real solar wind, a proton density of 8.2 cm$^{-3}$, a proton thermal velocity
of $35$ km/s, which corresponds to a thermal energy of 12.8 eV, and a $T_{\alpha}/T_{p}=4$ have been used. Moreover, a bulk
velocity of $400$ km/s has been taken into account. Then the distributions functions are reassigned to the field of view of CSW by
means of an averaging process. At this point, the counts that the sensors would detect in each energy-angular interval for each
species are computed \cite{Paschmann}:
\begin{equation}
\mbox{counts}_{i,j,k} = v_i^4 \cdot f_{i,j,k}\cdot T_{acc}\cdot G
\end{equation}
where $i,j,k$ refer to the energy, elevation and azimuth intervals, $v_i$ is the velocity in the interval, $f$ the distribution
function, $T_{acc}$ and $G$ the accumulation time and the geometric factor, respectively. $f$ is normalized in such a way
that its integral in velocity space gives the particle number density. Afterwards, we determine the corresponding moments.
\begin{figure*}
\includegraphics[width=7.4cm]{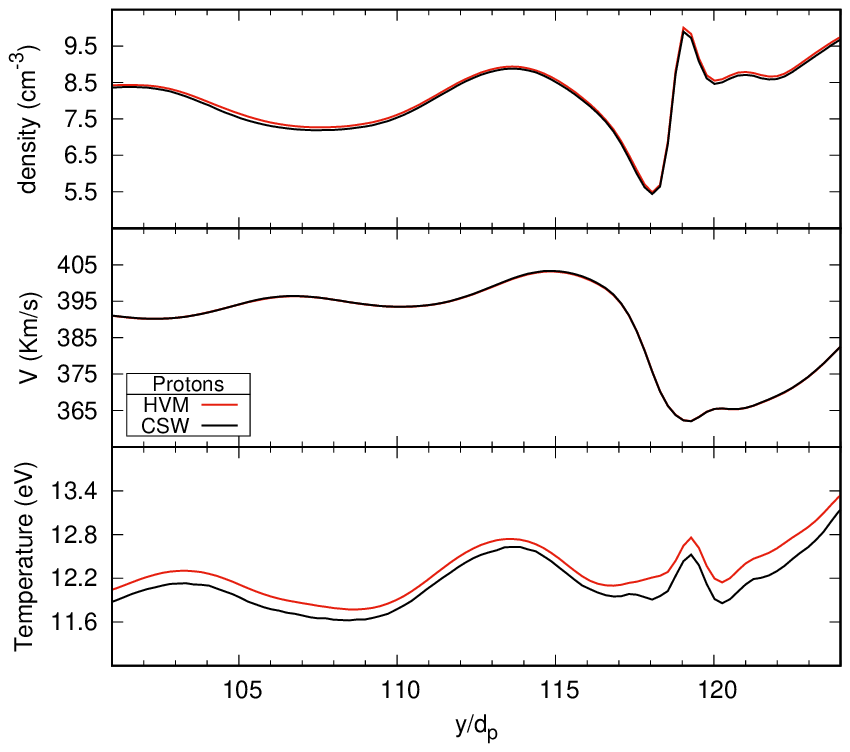}
\includegraphics[width=7.5cm]{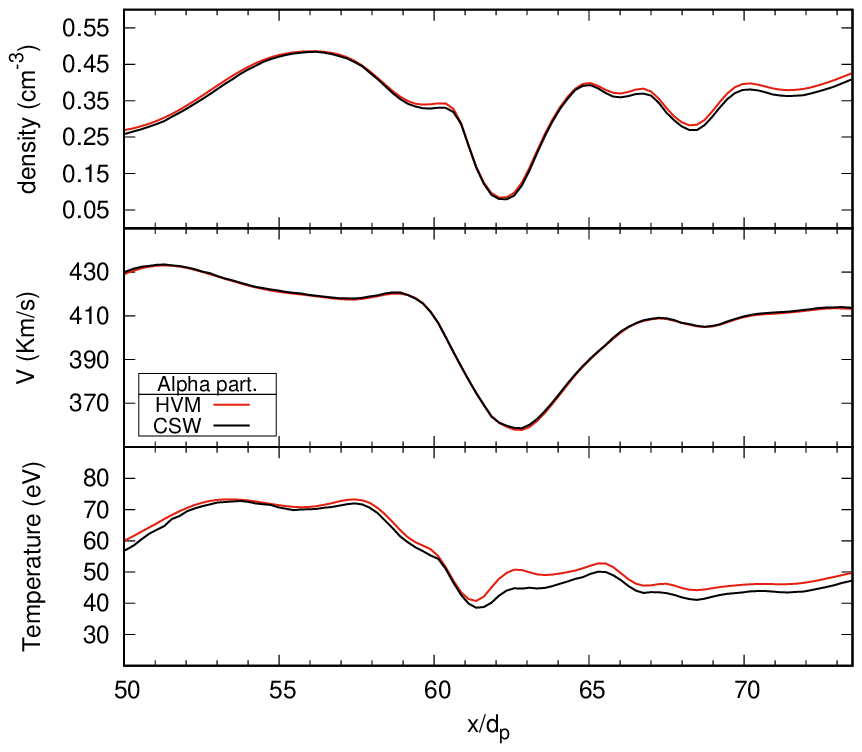}
\caption{HVM (red lines) and CSW (black lines) density, velocity and temperature, computed for protons (left) and alpha
particles (right) distributions along the $y$ and $x$ direction at $x_0\simeq 60 d_p$ and $y_0\simeq 115 d_p$, respectively.}
\lab{fig:moments}
\end{figure*}

In Figure \ref{fig:tophatsim} proton (top panels) and alpha (bottom panels) velocity distributions are shown at the
spatial point where $\epsilon$ is maximum, as recovered from the HVM simulations (left panels) and as sampled by CSW (right
panels). In particular, slices of the distribution functions in the $V_xV_z$ plane are presented.  

Looking at the protons, it can be seen that all the features that characterize the HVM distribution are revealed by CSW. In fact,
the ring-like modulations, that can be identified in the HVM distribution along the $V_x$ direction, are present also in the CSW
distribution, although they are less readily identifiable. Moreover, this distribution function spans only over one order of
magnitude and will be easily observed by CSW for regular or dense solar wind.

Looking at the alpha particles, it has to be noted that the alphas population appears in the CSW distribution at $\sqrt2$ times
the bulk velocity recovered from the simulation, since electrostatic analyzers like CSW detect particles with respect to their
energy-per-charge. Regarding the features characterizing the HVM and CSW alpha distributions, it can be seen that the shape of the
main alpha population, at about $(V_x,V_z)=(-340,30)$, is correctly reproduced. In fact, the elongation visible in the
HVM is clearly unveiled also in the CSW distribution, and also discernible is the faint widening on the left part of the core
population. Moreover, the faint blob at negative $V_z$ in the HVM distribution, is recognizable in the CSW distribution as well,
and its shape is well captured.

In Figure \ref{fig:moments} we show the HVM (red lines) and CSW (black lines) density, velocity and temperature, for protons
(left panel) and alpha particles (right panel), computed for distribution functions along the $y$ and $x$ direction at $x_0\simeq
60 d_p$ and $y_0\simeq 115 d_p$, respectively. Along these directions, the  maximum $\epsilon_p$ and $\epsilon_\alpha $ are
encountered. Results show that the velocity, and to a lesser extent the density, for HVM and CSW are extremely well in agreement.
Small differences can be noted between HVM and CSW temperature. The different behaviors of the temperature and lower order moments
was already observed in \cite{demarco16}. In general, larger errors are found between the HVM and CSW moments for alpha particles
than for protons. This can be explained in terms of the alpha particles lower counts statistics with respect to the protons.
\begin{figure*}
\includegraphics[width=15cm]{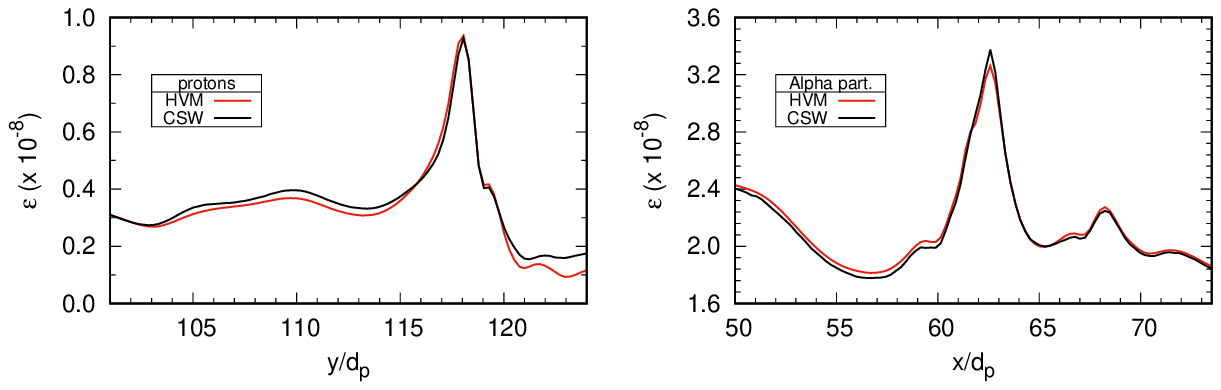}
\caption{The non Maxwellian measure $\epsilon_s$ computed for the HVM (red line) and CSW (black line) distributions of protons
(left panel) and alpha particles (right panel).}
\lab{fig:CSWepsilon}
\end{figure*}

Finally, the non Maxwellian measure $\epsilon_s$, computed for the HVM (red line line) and CSW (black line) distributions, is
presented in Figure \ref{fig:CSWepsilon}. The $\epsilon_s$ evolution is the same for the HVM and CSW distributions. In particular,
the two HVM $\epsilon_s$ peaks, of the protons and alphas cases, are well pronounced in the CSW  $\epsilon_s$ evolution as well.
The small difference between the peaks of the HVM and CSW $\epsilon_\alpha$ can probably be explained in terms of the alphas lower
counts statistics, like in the case of moments.  

It must be noted here that, starting from the HVM simulation, the moments for the alpha particles can be readily computed. In the
real case, although the alpha population can be easily identified in the CSW distribution functions, part of the protons and alpha
particles populations can overlap, especially when the plasma temperature is high, and the moments computation can be hard (we
recall that CSW selects particles according to their energy-per-charge). In such cases it is extremely useful to have a particle
instrument that is able to identify the different ion species. Such an instrument will be on board THOR: it is the Ion Mass
Spectrometer (IMS), a top-hat analyzer complemented with a time of flight section.

\section{Summary and conclusions}
We have employed a multi-component Eulerian hybrid Vlasov-Maxwell code to analyze the differential kinetic dynamics of the
dominant ions in the turbulent solar wind. We have performed simulations of decaying turbulence in a five-dimensional phase space,
varying relevant parameters, such as the level of turbulence and the plasma beta at equilibrium, in order to explore a large
portion of the parameter space and to compare our numerical results with direct solar wind measurements. As the numerical results
do not change qualitatively for the runs summarized in Table \ref{tab}, after showing that our simulations reproduce with a good
degree of realism the phenomenology recovered in solar wind observations, we restricted our analysis to a single simulation. 

We focused in particular on the departure of the plasma from thermodynamic equilibrium along the cascade and on the role of
kinetic effects coming into play when the cascading energy reaches the typical ion scales. As turbulence develops, we observed the
generation of ion temperature anisotropy with respect to the direction of the local magnetic field as well as a significant
heating for both ion species. Also, the non-Maxwellian index $\epsilon_s$, which is a measure of the deviations from Maxwellian of
each ion species, gets different from zero; at the same time, the generation of coherent structures (current sheets, vortices, 
etc.) of the typical width of few ion skin depths is observed in the contour map of the out of plane current density and of the
ion vorticity. Generation of temperature anisotropy, particle heating and deformation of the ion velocity distributions
clearly occur in the form of thin filaments at scales of the order of the ion kinetic scales.

A detailed statistical analysis of the simulation results provides evidence that the distribution of the numerical data is
constrained to the stable region with respect to the fire-hose and mirror stability thresholds, evaluated by taking into account
the multi-ion composition of the solar wind plasma. These results appear in good agreement with the observational evidences,
recently discussed in Ref. \cite{chen16}.

The generation of temperature anisotropy and the process of particle heating are significantly more efficient for the alpha
particles than for the protons and also the non-Maxwellian index gets larger values for the heavy ions. This means that not only a
differential heating occurs, but also a differential kinetic behavior is observed for the dominant ion species in the solar wind:
in fact, as shown in Figure \ref{fig:vdf}, the velocity distribution of the alpha particles appears much more distorted than that
of the protons. Interestingly, the process of differential ion heating takes place close to the peaks of the ion vorticity and in
spatial regions where the ion velocity distributions appear significantly distorted and far from thermodynamic equilibrium. 

In order to characterize the deviations from Maxwellian, we analyzed the ion velocity distribution both in its minimum variance
frame and in the local magnetic field frame, by defining the anisotropy and the gyrotropy indicators in both frames. The study of
these indicators reveals that, when kinetic effects are at work and $\epsilon_s$ gets large, the shape of the velocity
distributions of each ion species loses its symmetry, developing anisotropic and non-gyrotropic features. Moreover, close to the
peaks of $\epsilon_s$, sharp velocity gradients and complicated distortions are recovered.        

If, as suggested by our numerical results, temperature variations of the ion species are related to the kinetic dynamics at short
spatial scales, then the details of the velocity distribution, which keep the memory of the particle kinetic dynamics, are clearly
crucial ingredients for the understanding of the complex process of particle heating. In fact, as we discussed above, different
ion species, even though experiencing the same turbulent fields, exhibit a deeply different response, being heated and shaped
differentially. By feeding a virtual top-hat analyzer with the ion velocity distributions obtained from the kinetic
simulations discussed here, we have demonstrated that the energy and angular resolutions of the Cold Solar Wind instrument on
board the THOR mission should allow to obtain unprecedentedly high resolution measurements of both proton and alpha velocity
distributions and will therefore provide important information for unveiling the puzzling aspects of solar wind heating. The
results discussed in this paper support the idea that future space missions, designed to provide measurements of particle velocity
distributions and fields with very high resolution, will allow to gain relevant insights into the dynamics of the solar wind
plasma at kinetic scales.

\section*{Acknowledgements}
This work has been supported by the Agenzia Spaziale Italiana under the Contract No. ASI-INAF 2015-039-R.O ``Missione M4 di ESA:
Partecipazione Italiana alla fase di assessment della missione THOR''. The numerical simulations discussed in the present paper
have been performed on the Newton parallel machine at the Department of Physics at University of Calabria (Italy). Work at IRAP
was supported by CNRS and CNES. DS and SS aknowledge support from the Faculty of the European Space Astronomy Centre (ESAC).

\end{document}